\documentclass[12pt]{iopart}
\usepackage{fullpage}
\usepackage{amsthm}
\usepackage{amssymb}
\usepackage{graphicx}
\usepackage[export]{adjustbox}
\usepackage{iopams} 
\usepackage{color}
\usepackage{bm}
\usepackage{epsfig}
\usepackage{nicefrac}
\usepackage{lmodern}
\usepackage[T1]{fontenc}
\usepackage{textcomp}
\usepackage{subfig}
\usepackage{tabularx}

\begin{document}
\title{Optimization of a Superconducting Magnetic Energy Storage Device via a CPU-Efficient Semi-Analytical Simulation}

\author{I K Dimitrov$^{1,2}$, X Zhang$^{1}$, V F Solovyov$^{1}$, O Chubar$^{3}$ and Qiang Li$^{1}$}

\address{$^1$ Condensed Matter Physics and Materials Science Department, Brookhaven National Laboratory, Upton, New York 11973, USA}
\address{$^2$ System Evaluation Division, Institute for Defense Analyses, 4850 Mark Center Drive, Alexandria, VA 22311, USA}
\address{$^3$ Photon Sciences Directorate, Brookhaven National Laboratory, Upton, New York 11973, USA}

\eads{\mailto{idimitrov@bnl.gov}}

\vspace{2pc}

\begin{abstract}
Recent advances in second generation (YBCO) high temperature superconducting wire could potentially enable the design of super high performance energy storage devices that combine the high energy density of chemical storage with the high power of superconducting magnetic storage.  However, the high aspect ratio and considerable filament size of these wires requires the concomitant development of dedicated optimization methods that account for both the critical current density and ac losses in type II superconductors.  Here, we report on the novel application and results of a CPU-efficient semi-analytical computer code based on the \emph{Radia} 3D magnetostatics software package.  Our algorithm is used to simulate and optimize the energy density of a superconducting magnetic energy storage device model, based on design constraints, such as overall size and number of coils.  The rapid performance of the code is pivoted on analytical calculations of the magnetic field based on an efficient implementation of the Biot-Savart law for a large variety of 3D "base" geometries in the \emph{Radia} package.  The significantly-reduced CPU time and simple data input in conjunction with the consideration of realistic input variables, such as material-specific, temperature and magnetic field-dependent critical current densities have enabled the \emph{Radia}-based algorithm to outperform finite element approaches by a twenty-fold reduction in CPU time at the same accuracy levels.  Comparative simulations of MgB$_{2}$ and YBCO-based devices are performed at 4.2 K, and finally, calculations of the ac losses are computed in order to ascertain the realistic efficiency of the design configurations.            
\end{abstract}

\section{Introduction}

Superconducting magnetic energy storage (SMES) has been traditionally considered for power conditioning applications, where instantaneous high power can be delivered in a matter of milliseconds.  Among the advantages exclusively offered by superconductors we could mention: \emph{i}) higher energy storage efficiency ($\sim$95$\%$) than other existing energy storage systems \cite{Cheung}, and \emph{ii}) almost-immediate charge/discharge characteristics \cite{Badel,Ali}.  Certain applications, such as the electrical power grid and electromagnetic rail launchers are heavily-dependent on the efficiency and rapid charge/discharge response of a SMES \cite{Badel}.  Currently, it is one of a few superconductivity-based applications that is adaptable to the needs of the electric utilities markets \cite{Buckles}. 
 
Recently, a renewed interest in SMES technologies has been motivated by the search for means of improving the stability of the future power grid system, that would incorporate a large number of intermittent energy sources, such as wind and solar \cite{Kim,Jiang,Mikkonen}.  Historically, the concept of the SMES to a power system was originally developed in 1969 in order to balance the variations in the supply and demand of electricity \cite{Buckles}.  Low-temperature superconductors, such as Nb-Ti, have been successfully used for SMES, however, issues with reliability of 4.2 K cryogenics, efficiency of power electronics, and the relatively low energy density of Nb-based SMES have limited the applicability of the technology to a few cases, where electric power quality is at premium \cite{Huang1,Huang2,Rogers}. 

Since the magnetic energy stored in a SMES is proportional to ${B}^{2}$, doubling the operating field can quadruple the stored energy, an advantage, unmatched by other energy storage solutions.  In contrast to the superconducting transition temperature, ${T}_{c}$, and the upper critical field, ${H}_{c2}$, the current-carrying capability, measured by the critical current density, ${J}_{c}$, is governed by the vortex pinning strength, which depends fundamentally on the intrinsic properties of the superconducting state, and is determined by the ability of defects in superconductors to pin the vortices carrying magnetic flux.  

So far, second generation (2G) high temperature superconductor (HTS) YBa$_{2}$Cu$_{3}$O$_{7}$ (YBCO) has offered the greatest hope for implementation, since it exhibits all-around superior properties to all other classes of superconductors \cite{QL}, especially in light of the fact that it offers a possibility of operation at a temperature much higher than that of liquid helium \cite{Larbalestier}.  However, other promising candidates have emerged throughout the years as well, such as magnesium diboride, MgB$_{2}$ \cite{Nagamatsu} and the recently-discovered iron-based superconductors \cite{Kamihara}.  Even though MgB$_{2}$ and the Fe-based superconductors have typically lower ${T}_{c}$'s, ${H}_{c2}$'s and ${J}_{c}$'s compared to the HTS's, they exhibit much lower magnetic field anisotropies and are capable of being processed by a variety of methods \cite{QL}.    

Magnet design has become a process of paramount importance to SMES performance, necessitating better and faster analytical tools, since magnetic field calculations in and around a SMES demand sizeable CPU time and memory.  The Finite Element Method (FEM) has been the most common tool currently utilized for computations of magnetic fields in superconducting magnetic energy storage devices.  

In this report, we present an alternative approach towards building an algorithmic solution for simulating and optimizing a SMES device based on a selection of superconducting materials, such as second-generation YBCO tape or MgB$_{2}$.  This method is based on the \emph{Radia} software package, which is written in object-oriented C++, and interfaced to Mathematica via MathLink.  A substantial portion of algorithmic processing in the scope of this work is implemented in the Mathematica language.  Here, \emph{Radia} is used to assess the fields created by 3D volumes with constant current density, based on the Biot-Savart law.  Additionally, we present results on realistic ac loss assessment in a SMES, and suggest that for energies in the ${10}^{8}$ -- ${10}^{9}$ Joule range ac losses are tolerable and will not significantly impact SMES performance.

\section{Methodology}

\subsection{Overview}
In this section we present the analytical considerations behind our SMES design/optimization algorithm.  First, we begin by presenting an overview of the geometrical considerations for a realistic SMES.  Then, we describe \emph{Radia} -- an analytical tool used here as an alternative to finite element methods (FEM) calculations.  In the subsequent subsection we outline the computational considerations, and compare our initial outputs with those of a FEM.  In the final subsection, we revisit the topic of magnetic field scanning inside a coil as a necessary ingredient for a SMES optimization algorithm.    

\subsection{Geometrical Considerations}

Generally, there exist several types of geometrical arrangements for building a SMES \cite{Cheung,SMESWiki,Park}, but the toroidal design, offers the advantage of a reduced stray (perpendicular) magnetic field on the tape or wire, mostly confining the field inside the coil.  Such an arrangement is especially critical for second generation (2G) YBCO tape, since YBCO is marked by a high critical current density anisotropy, ${\gamma}_{m}$.  In the case of YBCO, the in-plane ${{J}_{c}}^{\parallel ab} ( B  = 17$ T) = 10.8 MA/cm$^{2}$, for example, can exceed the out of plane one ${{J}_{c}}^{\perp ab} ( B  = 17$ T) = 1.8 MA/cm$^{2}$ by roughly an order of magnitude at 4.2 K.    

\subsection{The Radia Software and Its Applications}    

The SMES simulation reported here has been built on top of the \emph{Radia} software package using \emph{Mathematica} by Wolfram Research.  The \emph{Radia} software package was designed by scientists at the European Synchrotron Radiation Facility (ESRF), for solving physical and technical problems one encounters during the development of insertion devices for synchrotron light sources.  However, it can also be used in different branches of physics, where efficient solutions of 3D boundary problems of magnetostatics are needed.  The \emph{Radia} package is essentially a 3D magnetostatics computer code optimized for undulators and wigglers.  The code has been extensively benchmarked with respect to a commercial finite element code \cite{Elleaume}.  All ESRF insertion devices built since 1992 have been designed using this code or its earlier versions.  A large number of predictions made by \emph{Radia} concerning the magnetic field and field integrals were verified on real insertion devices after manufacture.  Thus, one can design nearly any permanent magnet undulator or wiggler, including the central field as well as the extremities \cite{ESRF}.  

Creating and linking the objects properly is the first step in describing the magnetostatics problem.  Contrary to the FEM approach, \emph{Radia} does not mesh the vacuum (an example of a FEM geometrical segmentation and field generation is shown in Fig. 1) \cite{ESRF}, but rather, solves boundary magnetostatic problems with magnetized and current-carrying volumes, using the boundary integral approach.  The current-carrying elements can be straight or curved blocks \cite{Chubar}, and the planar boundary conditions are simulated via sets of space transformations, such as translations, rotations, or plane symmetry inversions \cite{Chubar}.  Applying transformations with a multiplicity can be understood as an efficient use of the symmetries in the model being solved.  This results in a minimum number of degrees of freedom, and therefore, dramatically reduces the memory requirements and CPU time needed to obtain a solution \cite{Elleaume}.  

\begin{figure}[ht]
\begin{center}
\includegraphics[scale = 0.62]{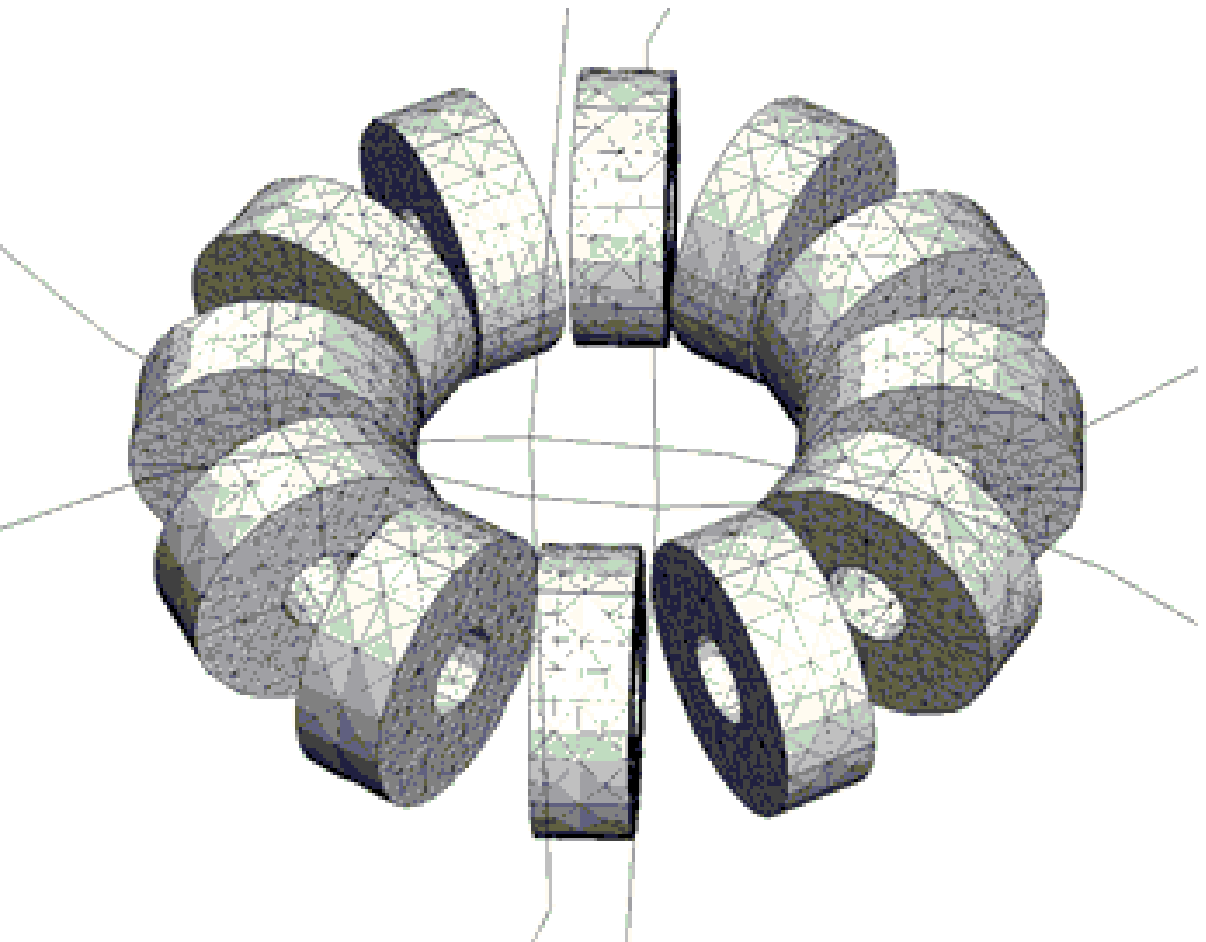}
\includegraphics[scale = 0.62]{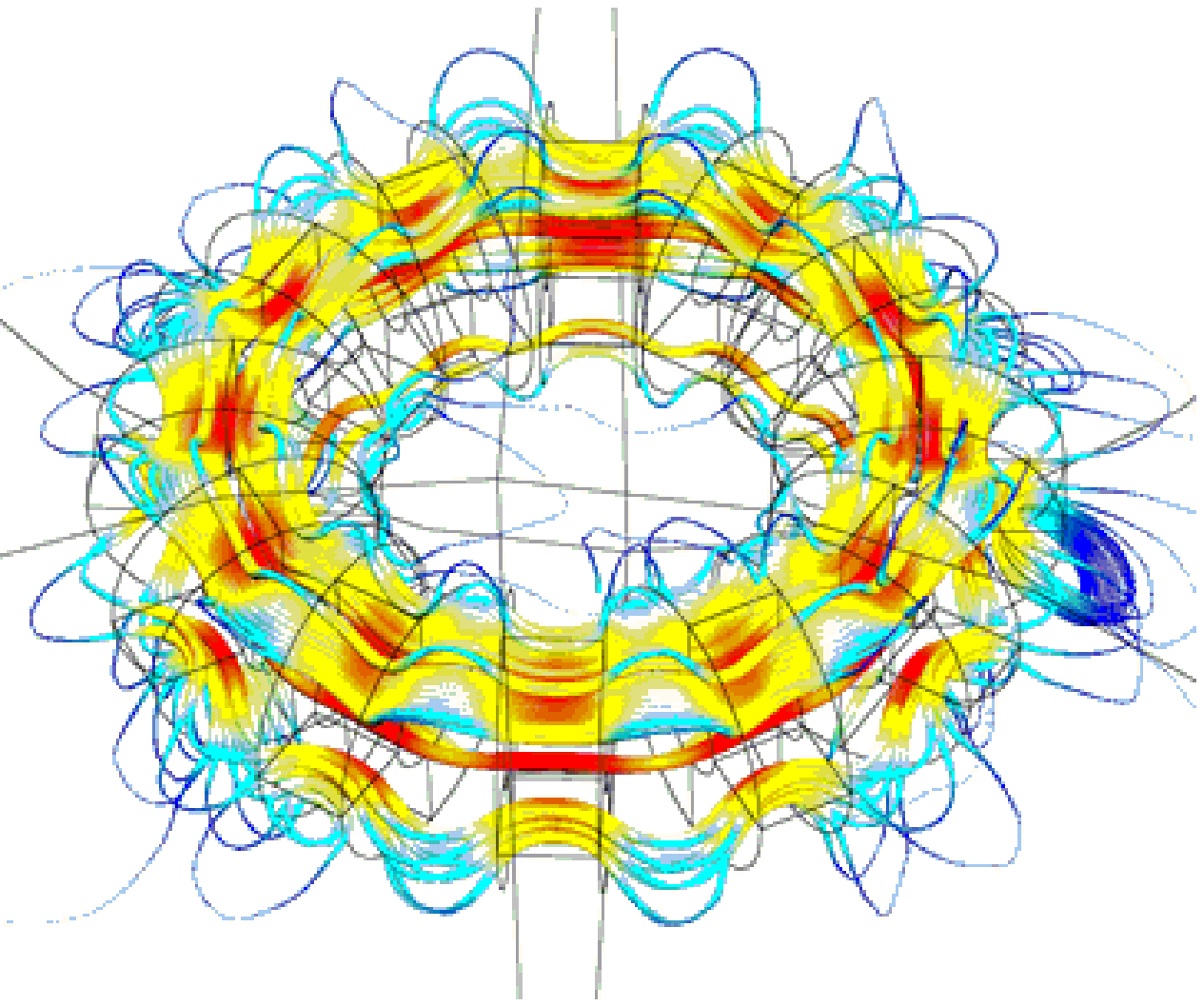}
\caption{A typical Finite Element Methods simulation of a SMES via COMSOL.  A simulation of the geometric dimensions and arrangement of the coils is shown on the left, whereas the resulting magnetic field distributions derived from a particular electrical current density are shown, juxtaposed onto the coil geometry, is shown on the right.}
\end{center}
\end{figure}

The reduction of the necessary elements for precisions comparable to FEM approaches (see Fig. 2) \cite{Sirois,Higashikawa} leads to a drastic reduction of the required CPU time (typically on the order of 20 times smaller).  This time efficiency is crucial when it comes to creating SMES optimization algorithms, where, as we will show in the Discussion section, for each point in configuration space one needs to carry the entire operation multiple times, from creating the entire SMES toroid with its associated coils, to computing the field distributions and stored energies, and calculating optimum design parameters.  

\begin{figure}[h]
\begin{center}
\includegraphics[scale = 0.25]{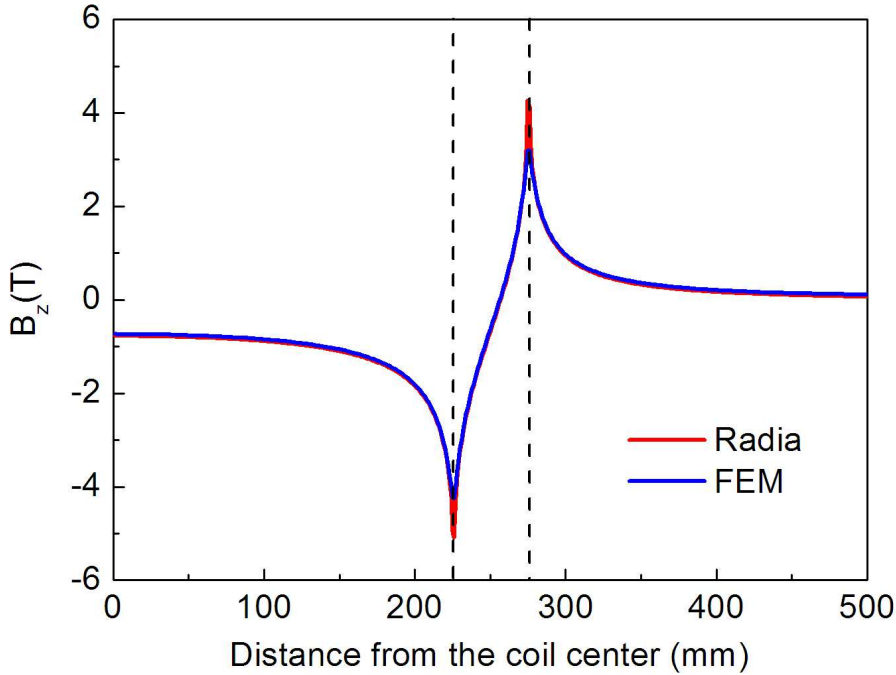}
\caption{Typical distribution of the magnetic field inside a superconducting coil calculated by \emph{i}) FEM (in blue), and \emph{ii}) \emph{Radia} (in red).  One of the issues with \emph{Radia} is the emergence of singularities at boundaries, as seen in the figure.  The two field profiles overlap everywhere except at the coil boundary, where \emph{Radia} unphysically diverges (due to the analytical approximation used).}
\end{center}
\end{figure}

In the current work, we employed \emph{Radia} to simulate and optimize a realistic SMES device, which takes into account the temperature and magnetic field dependent critical current density, ${J}_{c} \left( T, \, B \right)$, of the superconducting wire of choice used in the simulation (second generation YBCO tape or MgB$_{2}$).  The significant reduction in computation time was the primary motivation behind this endeavor.    

\subsection{Computational Procedure}

Our algorithm starts by specifying the actual geometry of the device, which is built according to given specifications, such as \emph{i}) coil radius, \emph{ii}) coil thickness, \emph{iii}) coil width, \emph{iv}) coil thickness, \emph{v}) toroidal radius, and \emph{vi}) number of coils.  The coil radius is defined as the mean of the inner and outer radii (with respect to the coil's axis of symmetry) of each coil, while the coil thickness is the difference between these two (see the 16 module toroidal system generated by \emph{Radia} in Fig. 3).  The large (toroidal) radius is the distance from the geometric center of the SMES to the geometric center of each individual coil, and the typical toroidal radia that we employed were on the order of 1 -- 2 meters.  A realistic toroidal-type SMES design must include the frame and conduction support bars, as well as insulating spacers between the coils. 

\begin{figure}[h]
\begin{center}
\includegraphics[scale = 0.17]{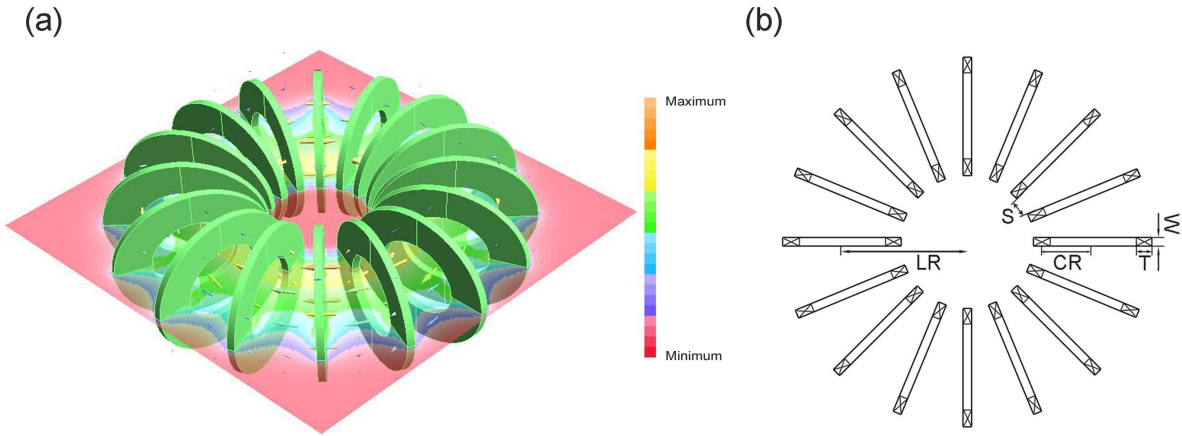}
\caption{(a) Structural concept of a 16-module toroidal system generated by \emph{Radia}, with b) design variables.  \emph{LR}, \emph{CR} and \emph{W} stand for the toroidal ("long") radius, coil radius and width, respectively.}
\end{center}
\end{figure} 

\emph{Radia} gives the freedom of choosing the degree of segmentation of each coil; the more segments are chosen, the closer the geometry is to an ideal cylindrical shape, but it will also result in longer computation time.  We "built" coils with 50 segments for the actual simulations.  A significant increase in the segmentation tends to lead to an undue increase in the computation time without the benefit of a significant increase in the precision.

In order to test the validity of \emph{Radia}-based simulations, we compared the outputs of our \emph{Radia}-generated SMES design with that of a FEM simulation by Lee \textit{et al.} \cite{Lee}.  We computed the maximum magnetic fields parallel and perpendicular to the test coil, as well as the total energy stored in the device using the same geometry and operating current density as used by Lee \textit{et al.}, and observed a discrepancy of less than 1$\%$ between the two methods.  The comparison between the two approaches is shown in Table 1.

\begin{table}[h]
\caption{\label{arttype}Comparison of \emph{Radia} and FEM-generated 2.5 MJ toroidal SMES configuration.}
\footnotesize\rm
\begin{tabular*}{\textwidth}{@{}l*{15}{@{\extracolsep{0pt plus12pt}}l}}
\br
Specification & In this work & From Lee \textit{et al.} \cite{Lee}\\
\mr
Operating Current Density (A/mm$^{2}$) & 194.5 & 194.5\\
Max parallel magnetic field (T) & 8.8 & 8.1\\
Max perpendicular magnetic field (T) & 1.12 & 1.16\\
Stored Energy (MJ) & 2.6 & 2.6\\
Total length of HTS conductor (km) & 19.58 & 19.55\\ 
\br
\end{tabular*}
\end{table}   

\subsection{Magnetic Field Distribution Computation}

The highest field inside a coil is the bottle-neck limiter to the amount of current density, $J$, that the wire can reasonably handle before it quenches and subsequently incapacitating the device.

We present magnetic field analysis results of a toroidal-type SMES magnet calculated via \emph{Radia} (see Fig. 4).  A characteristic magnetic flux density pattern of the center plane of the toroid is exhibited in Fig. 4a, while typical perpendicular, ${B}_{N} \left( x \right)$, and tangential, ${B}_{T} \left( x \right)$, flux density profiles along the radial direction of a single pancake coil are shown in Fig. 4b.  For this particular simulation the operating current of the magnet is taken to be 960 A, while the maximum perpendicular magnetic flux density, tangential magnetic flux density and stored energy of the simulated device obtained from the simulation are shown to be 1.01 T, 9.00 T and 2.68 MJ, respectively.  The discrepancies between the maxima of the \emph{i}) perpendicular magnetic flux density, and \emph{ii}) tangential magnetic flux density, obtained via FEM and \emph{Radia} were shown to be on the order of 2$\%$, confirming \emph{Radia}'s validity and reliability as an alternative tool to FEM for magnetic field analyses of SMES devices.

\begin{figure}[h]
\begin{center}
\includegraphics[scale = 0.13]{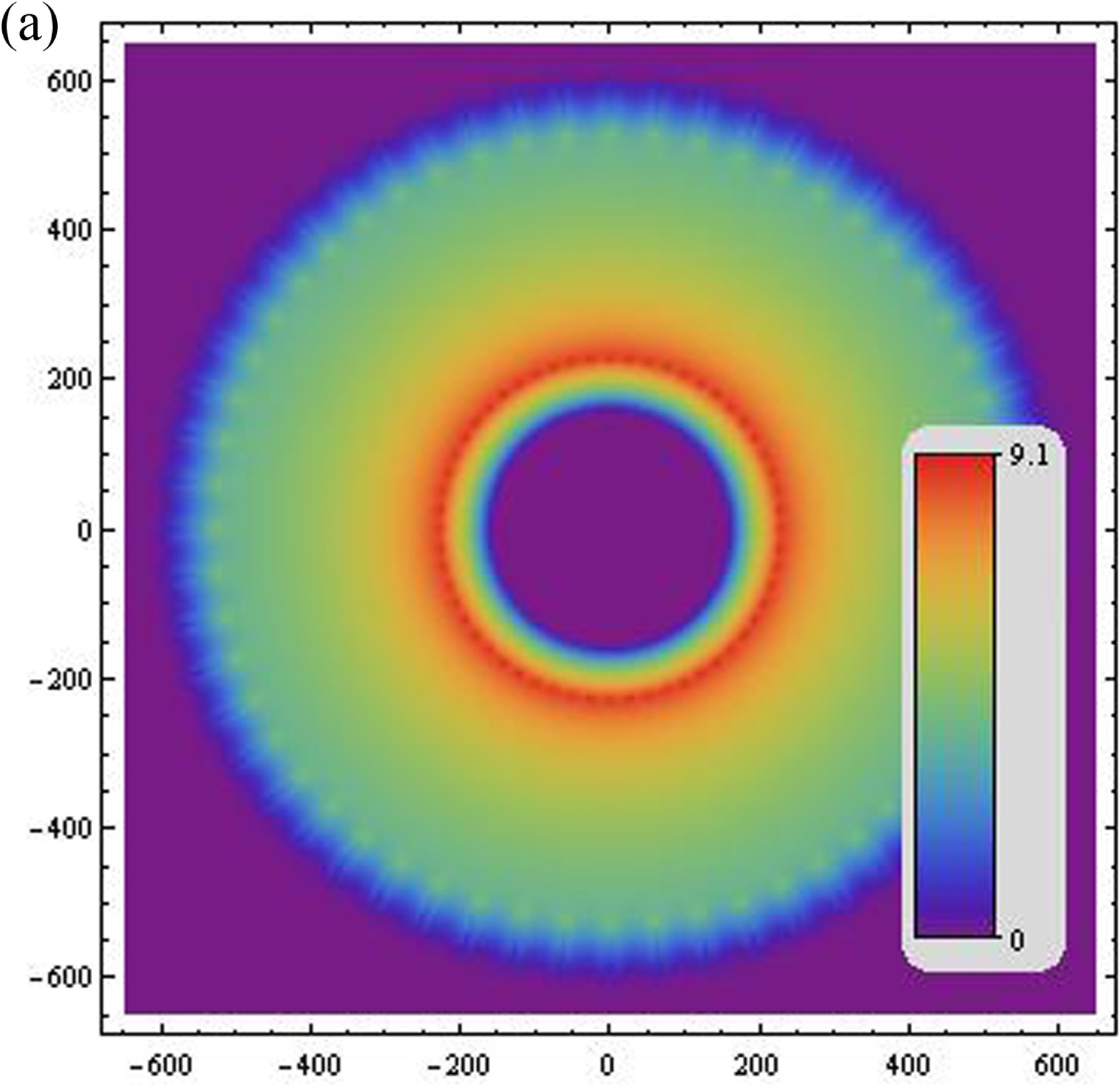}\hspace{1cm}
\includegraphics[scale = 0.13]{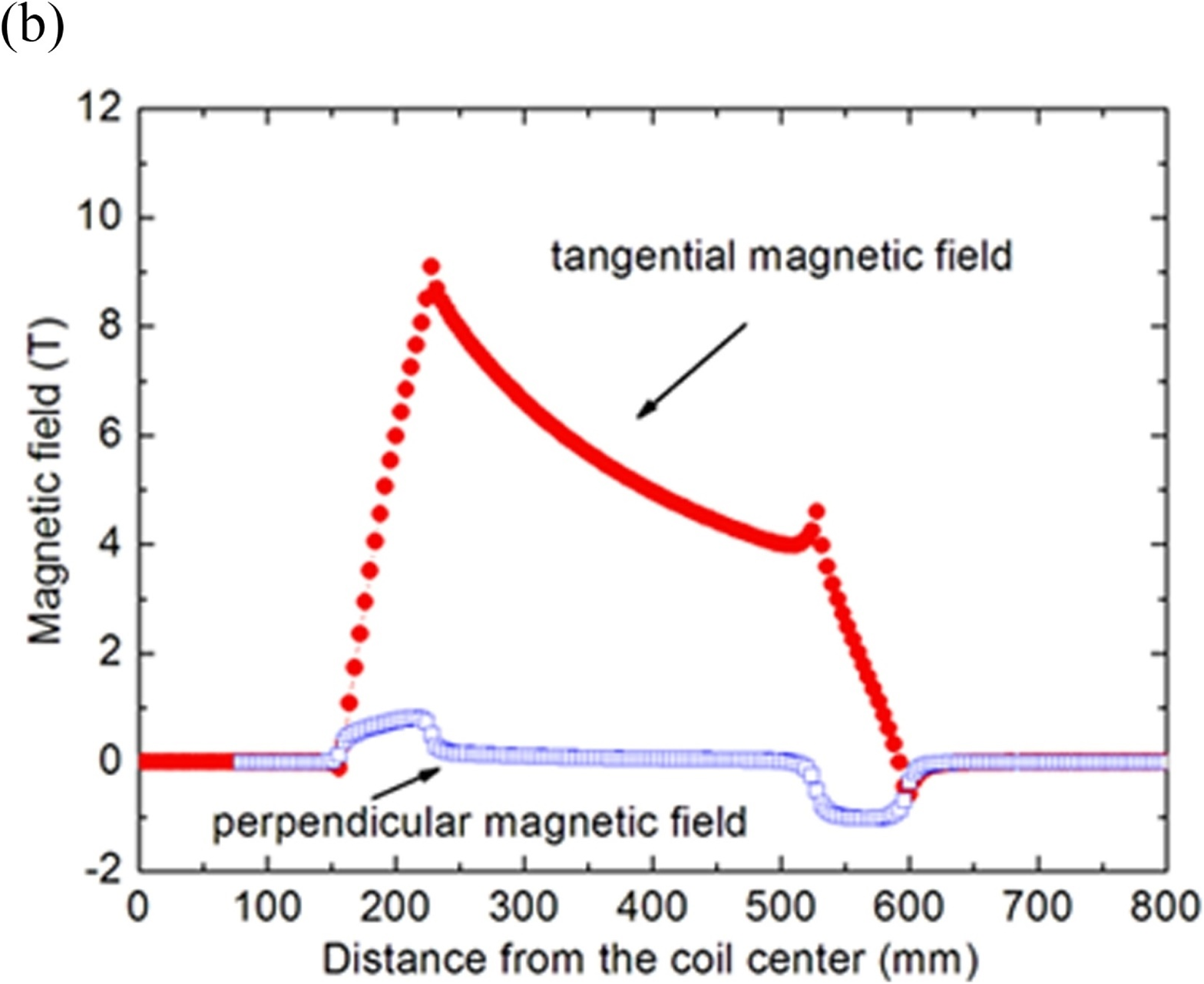}
\caption{(a) Magnetic flux density pattern of the center plane of toroidal-type SMES magnet; (b) Shown are the \emph{i}) perpendicular and \emph{ii}) tangential flux density analyses on a typical single pancake coil, in open gray and filled red circles, respectively.}
\end{center}
\end{figure} 

One of the conclusions one can immediately draw from the analysis of the field distribution, as evidenced in Fig. 4a, is that the field is mostly confined inside the coil.  Being able to confine the magnetic field inside the coil is important for a number of reasons: \emph{i}) large field gradients could be avoided from the coil edges (thus affecting the ${J}_{c} \left( B \right)$ of the wire), and \emph{ii}) stray fields are generally detrimental to electronic devices located in the vicinity of the magnet.  Thus, SMES designs with large numbers of coils may be preferred for as long as geometrical and critical current limitations are considered and met.    

Not surprisingly, the highest ${B}_{T}$'s are found at the inner rim of an individual coil, nearest the center of the torus along the \emph{z} = 0 plane, as evidenced in Fig. 4, \textit{i.e.}, ${B}_{T}$ does not exhibit a \emph{y}-axis mirror symmetry through the center of the coil, which is clearly due to the higher density of current-carrying elements closer to the torus center.  On the other hand, ${B}_{N}$ is shown to be several times smaller than ${B}_{T}$.  ${B}_{N}$ is supposed to vanish in the case of a perfect toroidal SMES with a continuum of coils.  However, there is always a finite ${B}_{N}$ for a SMES magnet built up with a discrete number of coils.  It should be noted that although ${B}_{N}$ is several times smaller than ${B}_{T}$, YBCO tapes are marked by a large anisotropy in the critical current density (${J}_{c} \left( {B}_{N} \right) \ll {J}_{c} \left( {B}_{T} \right)$), and therefore the maximum ${B}_{N}$ determines the current density cutoff that ought to be compared against the ${J}_{c} \left( B \right)$ used in SMES design.

\subsection{ac Loss Calculation}

The ac loss calculations are performed using a simplified algorithm.  The loss originating from the field over the penetration threshold is calculated using the accepted Brandt-Indenbom equation \cite{Brandt}, which takes into account the flat tape geometry of 2G wires as well as demagnetization effects.

\section{SMES Design}

\subsection{Optimization Algorithm Overall Flow}
Once the SMES geometry is specified (see Section 2.4), we feed the critical current density curves for \emph{i}) MgB$_{2}$ and \emph{ii}) 2G YBCO tape for different temperatures and field ranges into the code.  The algorithm fits all ${J}_{c} \left( B \right)$ curves for a given superconducting type at the specified temperature and the fitting parameters are stored for later use.  Subsequently, the magnetic field is scanned inside the coil and the maximal $B$, ${B}_{max}$, is utilized in the calculation of the minimal ${{J}_{c}}^{min}$, which would serve as the SMES design bottleneck.  The ratio of a trial value of the current density, ${J} \left( B \right)$, to ${{J}_{c}}^{min} \left( b \right)$ \cite{Ivo2}, is an indicator of the stability of the system, and is known as the "\emph{load factor}".  Depending on the value of the load factor, our algorithm chooses whether the thickness ought to increase or decrease in order for the requirement to be met.  Our load factor is set at 70$\%$ of ${{J}_{c}}^{min}$, so if $J > 0.7 \cdot{{J}_{c}}^{min}$, the algorithm increases the coil thickness and reduces $J$ by a proportionate amount since the total energy is proportional to the square of the current.  As long as the current is maintained at the same value, then so is the energy within a first order of approximation.  Analogously, if $J < 0.7 \cdot {{J}_{c}}^{min}$, the code "removes" turns (and proportionately increases $I$), until $J = 0.7 \cdot{{J}_{c}}^{min}$.  The increase of the coil thickness is essential in order to avoid a quench when the critical current density of the superconducting wire is exceeded.  The design flow of the entire optimization process is shown in Fig. 5

\begin{figure}[h]
\begin{center}
\includegraphics[scale = 0.55]{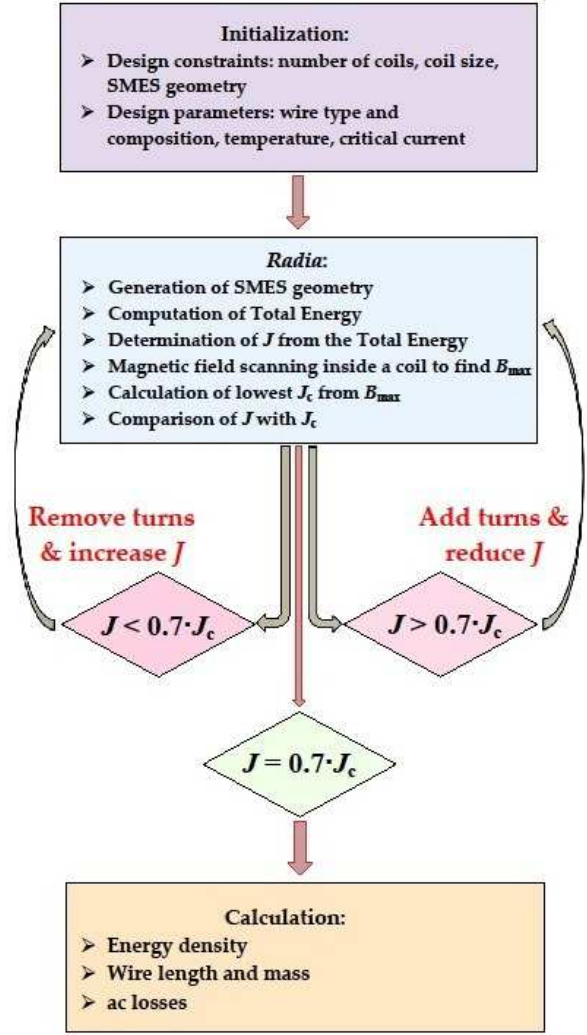}
\caption{Design process flow of SMES simulation and optimization routine.  The consecutive iteration generates the energy density (total stored energy / total wire mass) in $CR-W$ space.}
\end{center}
\end{figure}     

Once the SMES configuration is optimized, the algorithm performs the same operation as a function of \emph{i}) Coil Radius (\emph{CR}) and \emph{ii}) Coil Width (\emph{W}).  \emph{CR} is a continuous variable for the purposes of the simulation, and therefore it could be varied in steps of arbitrary size.  On the other hand, \emph{W} is a multiple of either the diameter of the wire (in the case of MgB$_{2}$, or the width of the tape (in the case of 2G wire).  Then, \emph{i}) the total energy, ${E}_{T}$, of the optimized configuration is computed, \emph{ii}) the ac losses are calculated as well, and \emph{iii}) ${E}_{T}$ is divided by the total wire mass of the SMES, resulting in the energy density, ${\rho}_{E}$.  

\subsection{Wires and Tapes}
Upon designing the SMES geometry based on a given toroidal radius and coil dimensions \cite{Ivo}, we specify the wire characteristics of two different conductors: \emph{i}) a MgB$_{2}$ wire which we assume in our calculation is a monofilamentary strand of MgB$_{2}$, manufactured by Hyper Tech Research, Inc., as discussed by Li \textit{et al.} \cite{Li} and \emph{ii}) standard second generation (2G) tape by American Superconductor, Inc. \cite{Malozemoff1,Malozemoff2,Rupich}.  The rapid advances in applied superconductivity, such as the successful implementation of pulsed-laser deposition techniques to grow iron-chalcogenide superconducting film on metal substrates \cite{QL,Mazin,Gurevich,Dimitrov,Liu,WDSi1,WDSi2} will hopefully lead to the development of commercial Fe-based superconducting wire in the near future, as well.

The MgB$_{2}$ wire our simulation is based upon was manufactured via an internal magnesium diffusion (IMD) method.  The particular wire included in our simulation was researched and discussed by Li \emph{et al.} \cite{Li}.  It consists of an annulus of MgB$_{2}$ enclosed in a Nb chemical barrier and an outer monel sheath \cite{Ivo3}.  The other choice for superconducting wire used in our simulation, 2G YBCO tape, is standard American Superconductor second-generation YBCO tape \cite{Ivo4}.

\subsection{Critical Current Densities}

Subsequently, we specify the critical current densities of second generation (2G) YBCO tape \cite{Malozemoff1,Malozemoff2,Rupich} and MgB$_{2}$ wire \cite{Li} at a spread of temperatures.  In the case of MgB$_{2}$ we use data for ${J}_{c}$ versus $B$ at $T$ = 4.2 K, 10 K, 15 K, 20 K, and 25 K \cite{Li}, and the 2G ${J}_{c} \left( B \right)$ data is available to us at $T$ = 4.2 K, 20 K, 30 K, 40 K, 50 K, 65 K and 77 K \cite{Xu, Braccini}.

${J}_{c} \left( B \right)$ decays as a function of the magnetic field.  Empirically, ${J}_{c} \left( B \right)$ could be fitted via a stretched exponential or a power-law form (depending on the model of choice).  Therefore, it is very important to parse the data according to the magnetic field range.  When stretched exponential or power-law experimental data is fitted with the appropriate function, the fitting algorithm typically follows a ${\chi}^{2}$ approach, \emph{i.e.}, the effective fit is that which reduces the sum of the squares of the differences between the fit and the data points.  Consequently, the part of the function most susceptible to errors will be the tail, since, in our case, small changes in the shape of the curve would lead to big changes in the critical current density.  Therefore, for ${J}_{c} \left( T, B \right)$ we produced a series of curves relevant in different field ranges.  For $B \leq 1$ T we used the entire magnetic field range.  For 1 T$< B \leq 2$ T, we used ${J}_{c} \left( B \right)$ above 1 T, for 2 T$< B \leq 3$ T we used the full critical current set above 2 T, and so on.  If the highest scanned $B$ exceeds the highest available experimental data $B$, the whole process is automatically aborted.  We noticed a dramatic improvement in the field consistency, which is most clearly manifest in the relatively smooth energy density surfaces, exhibited in the Results and Discussion section. 

Analytical fits of double-bending functions were used to obtain functional forms for ${J}_{c}$'s at different temperatures for both 2G YBCO and MgB$_{2}$ wires.  In principle, double-bending functions have been shown to be very effective in fitting the front and tail ends of ${J}_{c} \left( B \right)$, as if the two data curvatures are distinguishable \cite{Schwerg}, very much in the spirit of the data parsing used in the current research effort.  However, being able to discern between two functional forms needed for a fit would require a much greater density of ${J}_{c} \left( B \right)$ data points, which would be a necessary requirement for a realistic SMES engineering.   

The correct assessment of the ${J}_{c} \left( B \right)$ would be subsequently used in order to "thicken" or "thin" the coil by the algorithm in order to achieve optimal coil thickness.

\subsection{Assessing ${B}_{max}$}

In Section 2.5, we discussed the methodology and importance of field scanning inside a typical SMES coil.  However, there are a number of prospective SMES applications which require stored energies $\sim {10}^{7}$ -- ${10}^{9}$ J, and most of our efforts have been focused predominantly on that energy range, particularly when considering overall dimensions on the 1 -- 2 meter scale.  The dimensional considerations of such a device are very important, especially if it is to be transportable and/or integrated with other technological instruments.   

In order to devise the algorithm that assesses the maximal internal magnetic field, we consider two ${10}^{7}$ Joule SMES design cases at 4.2 K: \emph{i}) a SMES built of MgB$_{2}$ wire, and \emph{ii}) a SMES built of 2G tape, both of which contain 20 turns, Long Radius, \emph{LR}, = 1000 mm, coil radius, \emph{CR}, = 200 mm, coil width, \emph{W}, = 96 mm, coil thickness, \emph{T}, = 100 mm.  Then, we scan the magnetic field of the MgB$_{2}$ SMES coil transversely, going through the middle of the coil and assessing the absolute value of $B$ in the MgB$_{2}$ case, where the field is expected to be the highest (see Inset of Fig. 6a).  In the case of 2G wire, we scan along the edge of the coil, as shown in the Inset of Fig. 6b, and assess only the perpendicular $B$ component, since this is the application bottleneck, as discussed in Section 2.5.    

\begin{figure}[h]
\begin{center}
\includegraphics[scale = 0.42]{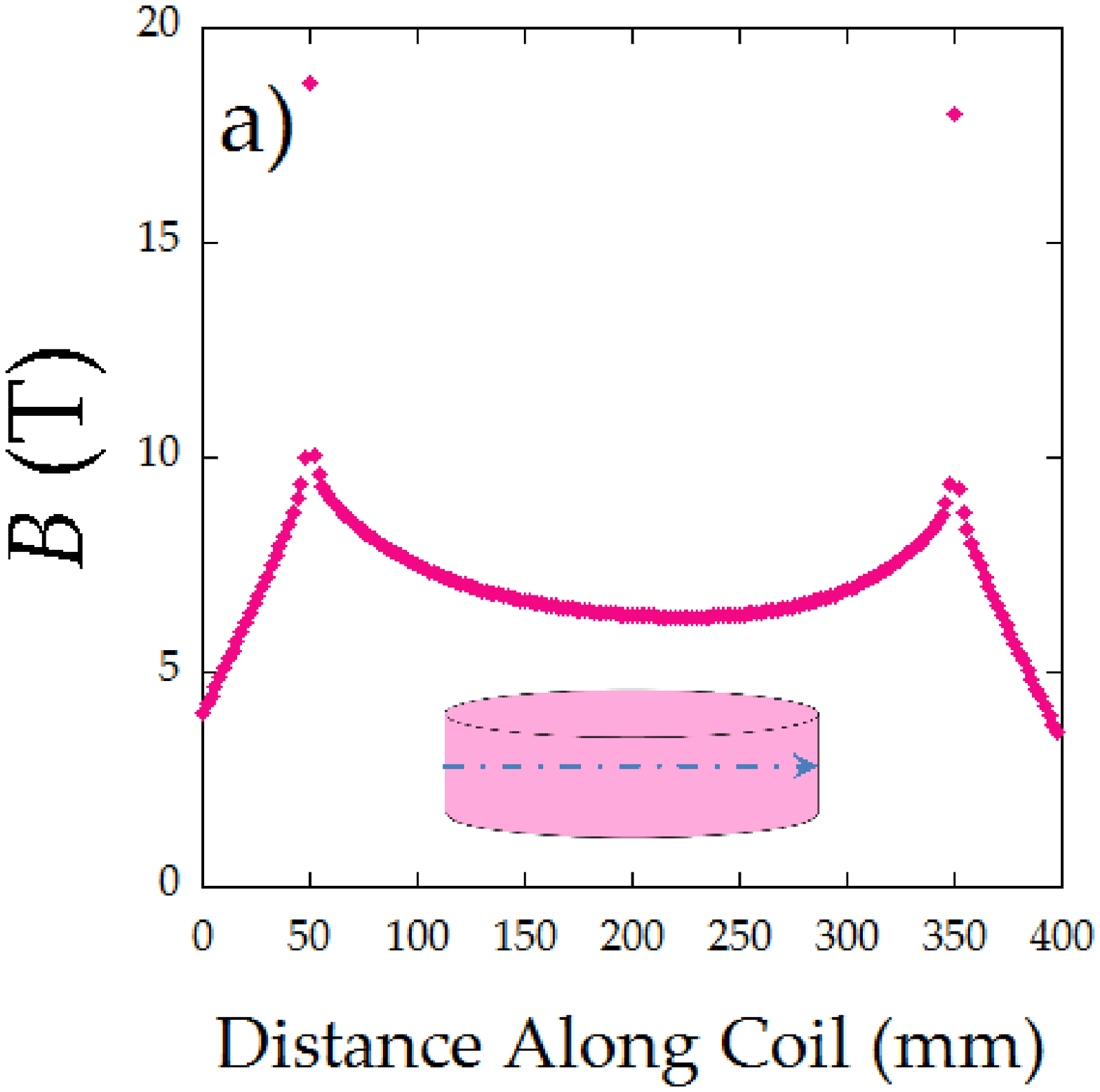}\hspace{1cm}
\includegraphics[scale = 0.42]{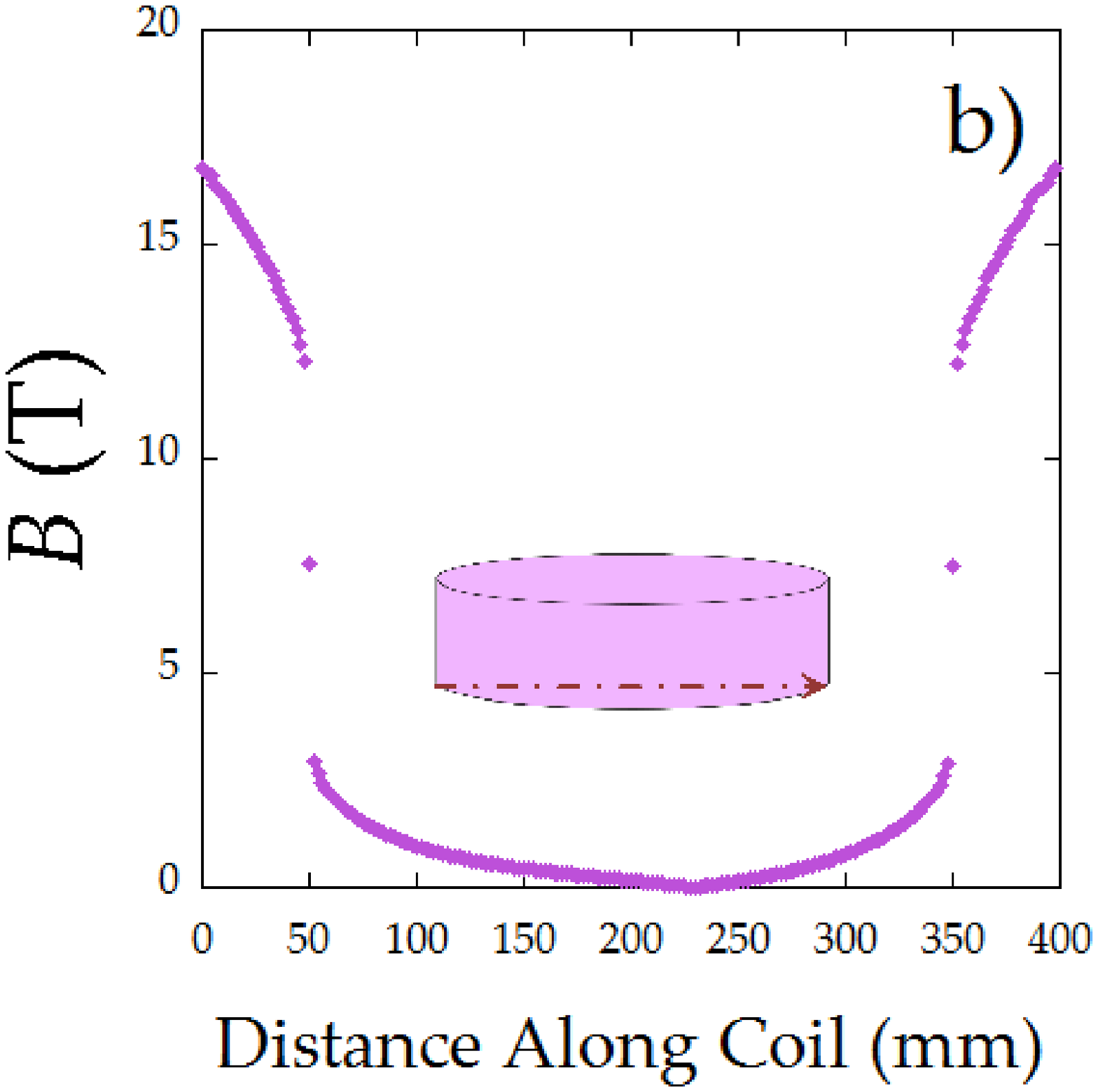}
\caption{Transverse magnetic field scans through a) MgB$_{2}$ and b) YBCO SMES coils.  The directions of scanning are shown in the insets of each figure.  The highest total $B$ fields are expected in the middle of a coil, and thus we scan across the middle of the coil for MgB$_{2}$.  However, in the case of YBCO, only ${B}_{N}$ is assessed (see Section 2.5), along the edge, where the field "leaks out" between the coils and thus, the highest ${B}_{N}$ is expected.  Note the singular points at 50 mm and 350 mm from the inner edge of the coil -- a \emph{Radia} artifact, which need to be removed from the analysis.}
\end{center}
\end{figure} 

As one can see from Fig. 6a, \emph{Radia} tends to produce an unphysical singularity in the vicinity of ${B}_{max}$, as evidenced in the points at 50 mm and 350 mm from the inner edge of the coil at the stored energy used in the simulation.  Choosing any of these two points ("artifacts" on the conductor boundary) could lead to erroneous ${J}_{c} \left( B \right)$ results.  The excessively high field at the conductor boundary does not disappear as we change the level of segmentation of each coil.  \emph{Radia} assumes a constant ${J}_{c}$ over the entire coil and does not perform any "relaxation" with respect to the current density/field in different areas of the conductor, suggesting this could lead to an excessively high field at the edge, and pointing to an area of future improvement.  Thus, instead of picking the maximal point from the scan (the weakest point), we identify its location, and fit the data points on either side of the singularity with a tenth order polynomial and get the mean of the two functions at the intersection (maximal point) location.  We get ${B}_{max}$ = 10.68 T using this algorithm, which is consistent with the cusp at 50 mm from the inner edge of the coil.  This procedure has led to smooth analytic energy density surfaces, to be discussed later.   

Analogously, in order to avoid any \emph{Radia}-generated artifacts, we have imposed a 90$\%$ cutoff from the scanned ${B}_{max}$ in the YBCO case, \emph{i.e.}, only field values $\leq 90 \%$ of ${B}_{max}$ are considered.  The truncated data is subsequently fitted with a tenth order polynomial and the maximal value is extracted from this fit.

The resulting ${B}_{max}$ were used to calculate the (lowest) critical current densities found in a single SMES coil for a given geometry and energy storage requirement, as outlined in the previous subsection.  Every time an input parameter changes, the code would scan for ${B}_{max}$ \emph{de novo}.

\subsection{ac Losses Calculation}

After the assessment of ${B}_{max}$ and the evaluation of its corresponding ${J}_{c} \left( {B}_{max} \right)$, the algorithm executes a comparative loop of the trial $J$, which (along with the coil thickness) are varied until the loading factor requirement is met.  Finally, after the optimal configuration is derived and ${E}_{T}$ recomputed, we proceed to calculate the associated ac losses.

The total ac loss per cycle in a YBCO superconductor, ${E}_{L}$, is given by \cite{Brandt}:
\begin{equation}
{E}_{L} = {B}^{\perp ab} \cdot TW \cdot {I}_{c} \left( {B}^{\perp ab} \right)
\end{equation}
, where ${B}^{\perp ab}$, \emph{TW} and ${I}_{c}$ stand for the magnetic field perpendicular to the tape, the thickness of the tape and the critical current.  The same analysis could be undertaken for MgB$_{2}$, but due to the essentially insignificant anisotropy, ${B}^{\perp ab}$ could be essentially replaced with the total $B$.

For the purposes of the calculation, we took a coil and segmented it in small pieces and scanned $B$ inside of each one, in order to obtain ${J}_{c} \left( B \right)$.  Because of the symmetry of the coil, we analyzed only a quarter of the test coil, focusing on azimuthal angles in the ${0}^{\circ} \leq \theta \leq {180}^{\circ}$ ($\theta$ going from ${0}^{\circ}$ to ${180}^{\circ}$ in ten steps).  In the dimension along the coil's symmetry axis, \emph{i.e.}, the $z$-axis, we scanned $B$ from the middle of the coil to the coil to the edge, \emph{i.e.}, from $z$ = 0 to $z$ = $W/2$ using the wire thickness as the step size in the MgB$_{2}$ case, while in the YBCO case we arbitrarily broke the $z$ scan into 5 pieces.  The radial axis, $r$, sedimentation took place from the coil inner radius to its outer radius using the wire/coil width of the MgB$_{2}$ wire and the 2G tape as the natural step size for the assessment of $B$ inside of each small cube.  Subsequently, we evaluated ${J}_{c} \left( B \right)$ for each cube in every shell (constant $r$ surface) of the coil, averaged ${J}_{c} \left( B \right) \cdot B$ for every shell and multiplied this average by the differential volume of the shell, getting the areal energy term ${E}_{A}$.  The ${E}_{A}$'s of all shells are added to obtain ${{E}_{A}}^{Total}$ before finally, we calculate the ${E}_{L}$ for the MgB$_{2}$ case:
\begin{equation}
{{E}_{L}}^{MgB_{2}} = \frac{ \left( 4 \pi N \right) \cdot \left( {{E}_{A}}^{Total} \right) \cdot  {\left( \nicefrac{TM}{2} \right)}^{3} }{{V}_{T}}
\end{equation}
, where \emph{N}, \emph{TM} and ${V}_{T}$ stand for the number of coils, the thickness of the monocore wire and the total volume displaced by a coil.  Note, that the \emph{4} in the numerator is necessary to account for the fact that we actually scan over the volume of a quarter of a coil.  Analogously, in the case of 2G YBCO tape, the total ac loss per cycle is calculated according to:
\begin{equation}
{{E}_{L}}^{YBCO} = \frac{ \left( 4 N \right) \cdot \left( {{E}_{A}}^{Total} \right) \cdot  {\left( TW \right)}^{2} \cdot \left( TT \right)}{{V}_{T}}
\end{equation}     
, where \emph{TT} stands for the thickness of the 2G tape.

\section{Results and Discussion}

Simulations were run for both 2G tape and MgB$_{2}$ wire at 4.2 K.  In the case of the (2G) YBCO device the simulation was run for a SMES comprising 20 coils, initial coil radius, ${CR}_{0}$, = 400 mm, initial coil width, ${W}_{0}$, = 96 mm, initial coil thickness, ${T}_{0}$, = 100 mm, and constant \textit{LR} = 1500 mm.  In the case of the MgB$_{2}$ wire-based SMES 32 coils were considered, along with ${CR}_{0}$ = 400, ${W}_{0}$, = 96 mm, ${T}_{0}$, = 100 mm and \textit{LR} = 2000 mm.  For both simulations the number of segments was chosen to be 50, and the number of points considered in the configuration space of the project (the $CR - W$ plane) was fixed at 1,500.   

For both simulations ${CR}$ was increased from ${CR}_{0}$ = 400 mm to its final value, $CR$ = 700 mm, in steps of 2 mm.  The coil widths for the 2G and MgB$_{2}$-based devices, however, were varied differently, since in the former case the natural step size the thickness of the tape (12 mm) while in the latter case, it is the thickness of the monocore strand.  Therefore, for the YBCO-based SMES, $W$ was varied in steps of 12 mm, while in the MgB$_{2}$ case, we arbitrarily picked the step size to be 12$\times$0.933 mm (monocore thickness = 0.933 mm).  Larger coil radii and widths were purposely omitted from the simulation, since we noticed that at larger values of those, the coils would start to physically "overlap", leading to unphysical solutions.  

Both sets of devices were simulated under the initial provision that they would store ${10}^{8}$ Joules of energy.  The energy settings were subject to small changes, subordinating a stringent energy requirement to the objective of creating an optimized (70$\%$ load factor) SMES.  The optimization was performed at every point in the specified segment of $CR - W$ space, and the \emph{i}) energy, \emph{ii}) energy density, \emph{iii}) \emph{T}, \emph{iv}) \emph{CR}, \emph{v})\emph{W}, \emph{vi}) ac losses, \emph{vii}) total SMES mass, \emph{M}, and \emph{viii}) total wire length needed, \emph{L}, were assessed at each of the 1,500 points (see Fig. 7).  

\begin{figure}[h]
\begin{center}
\includegraphics[scale = 0.14]{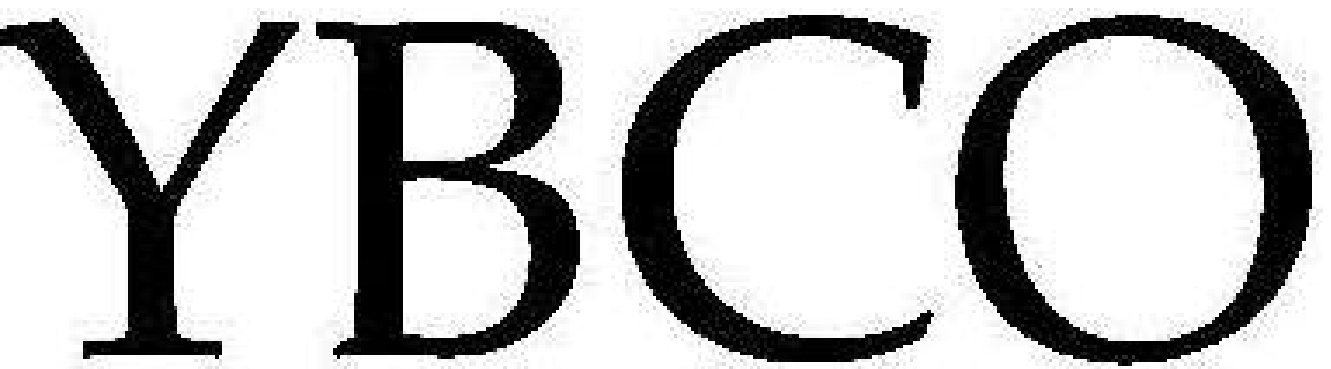}\hspace{6cm}
\includegraphics[scale = 0.14]{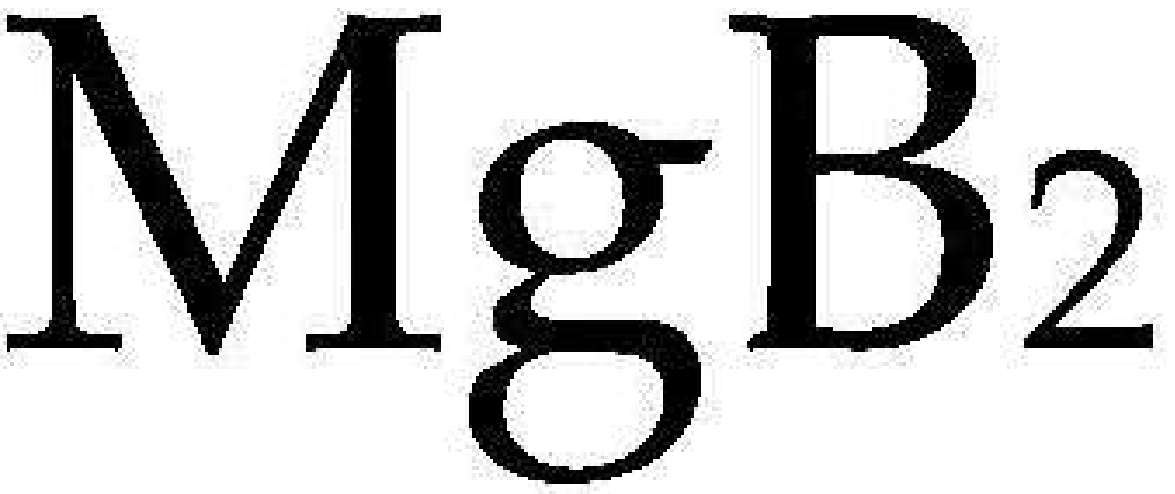}\\
\includegraphics[scale = 0.14]{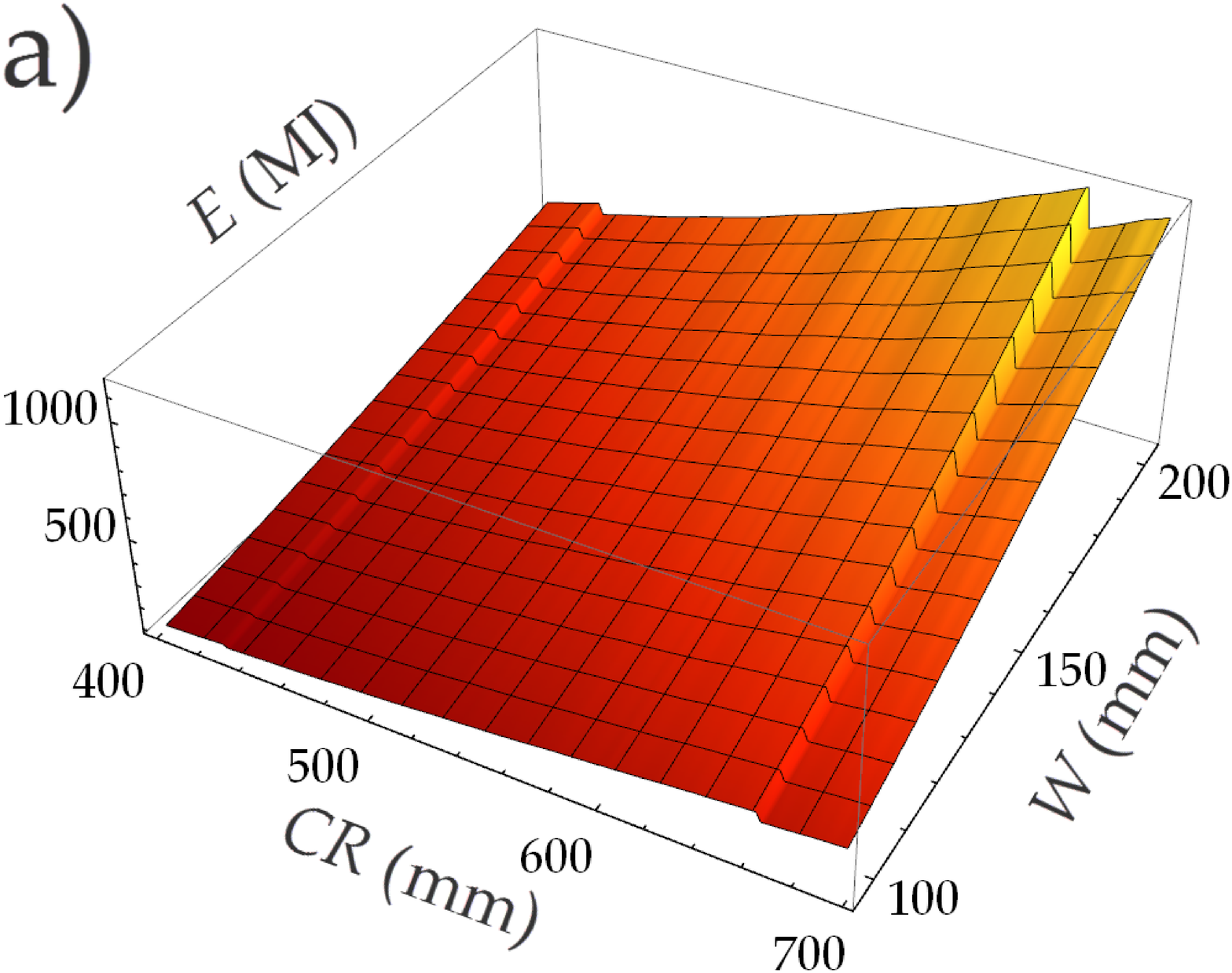}\hspace{2cm}
\includegraphics[scale = 0.14]{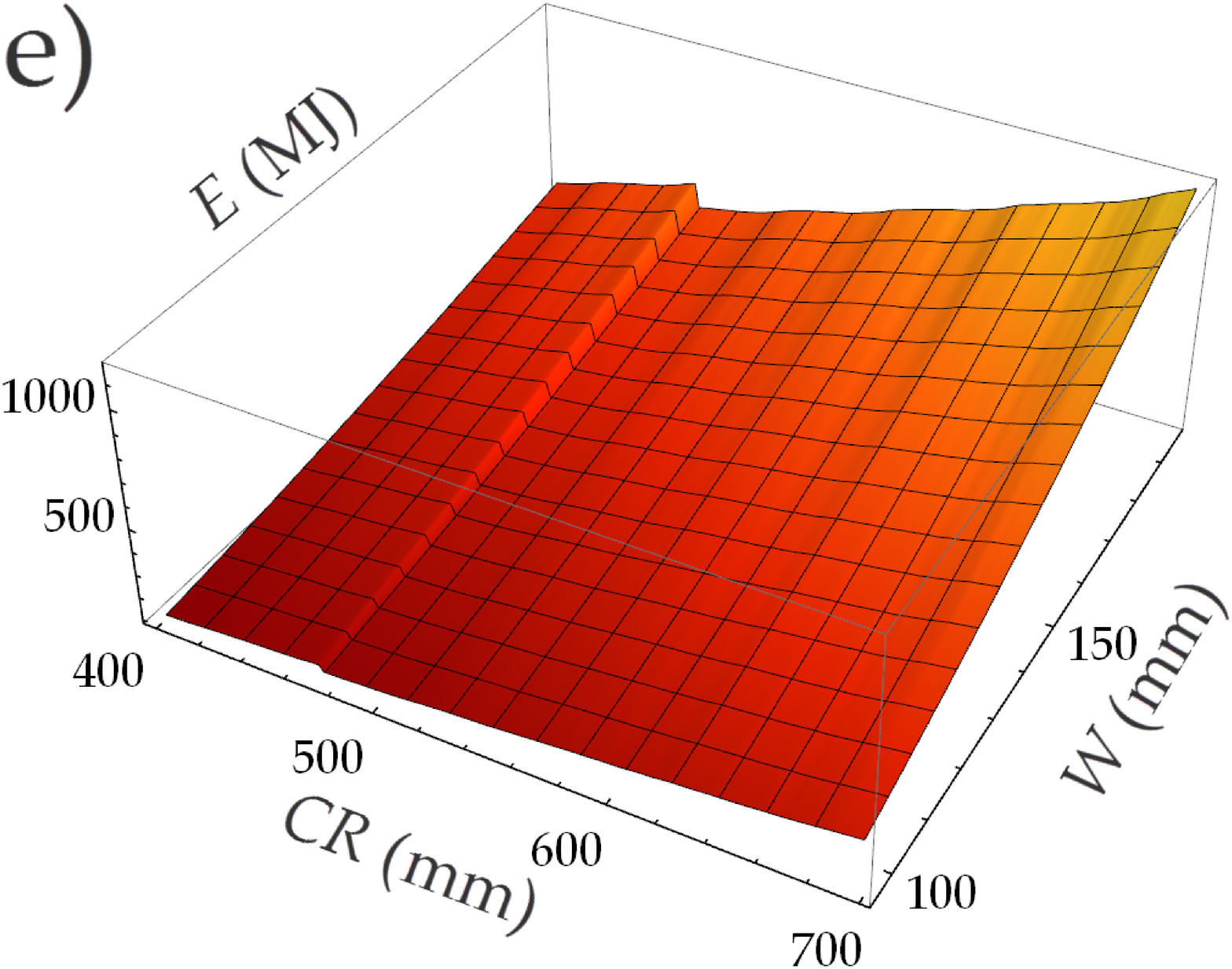}
\vspace*{0.2cm}
\includegraphics[scale = 0.14]{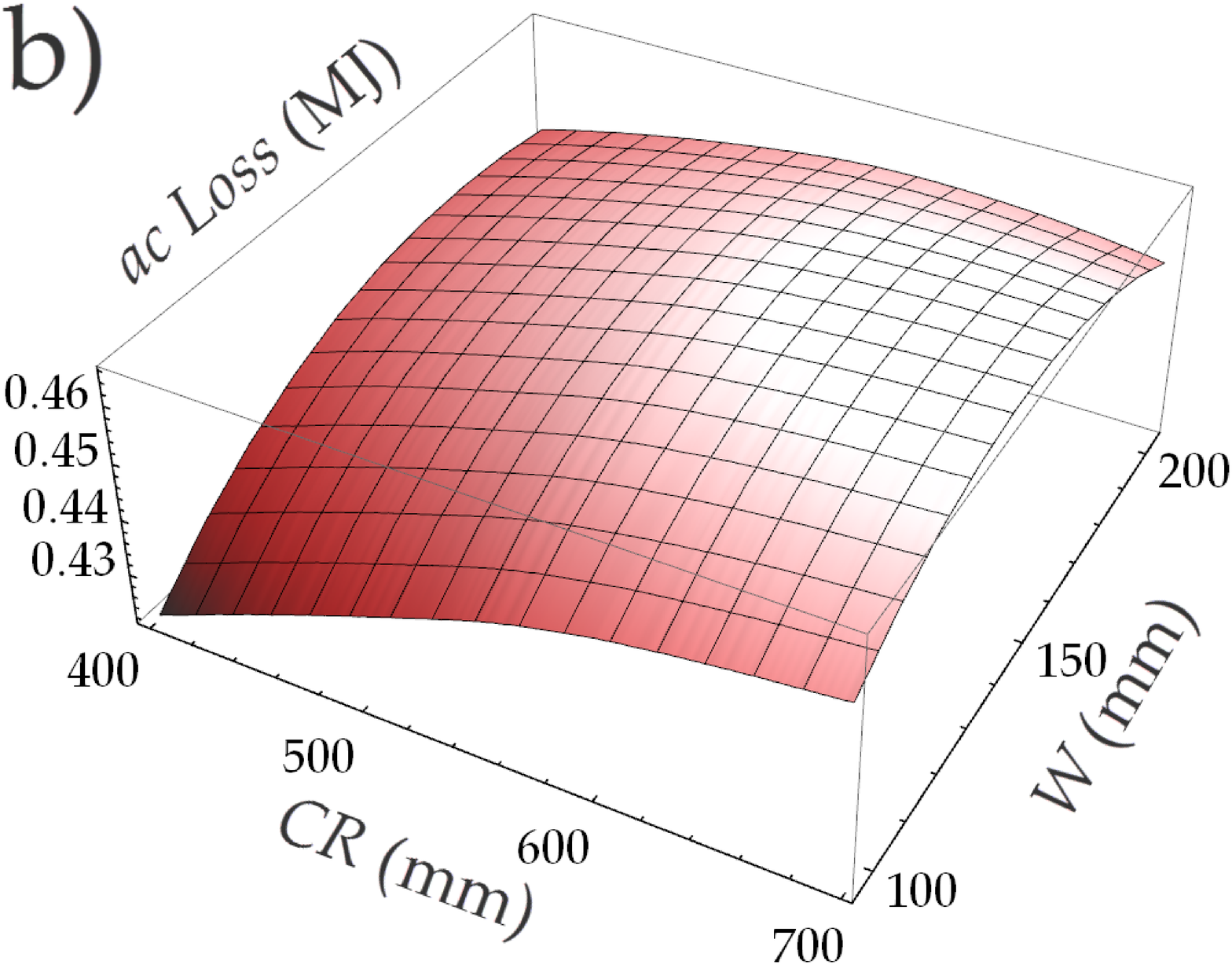}\hspace{2cm}
\includegraphics[scale = 0.14]{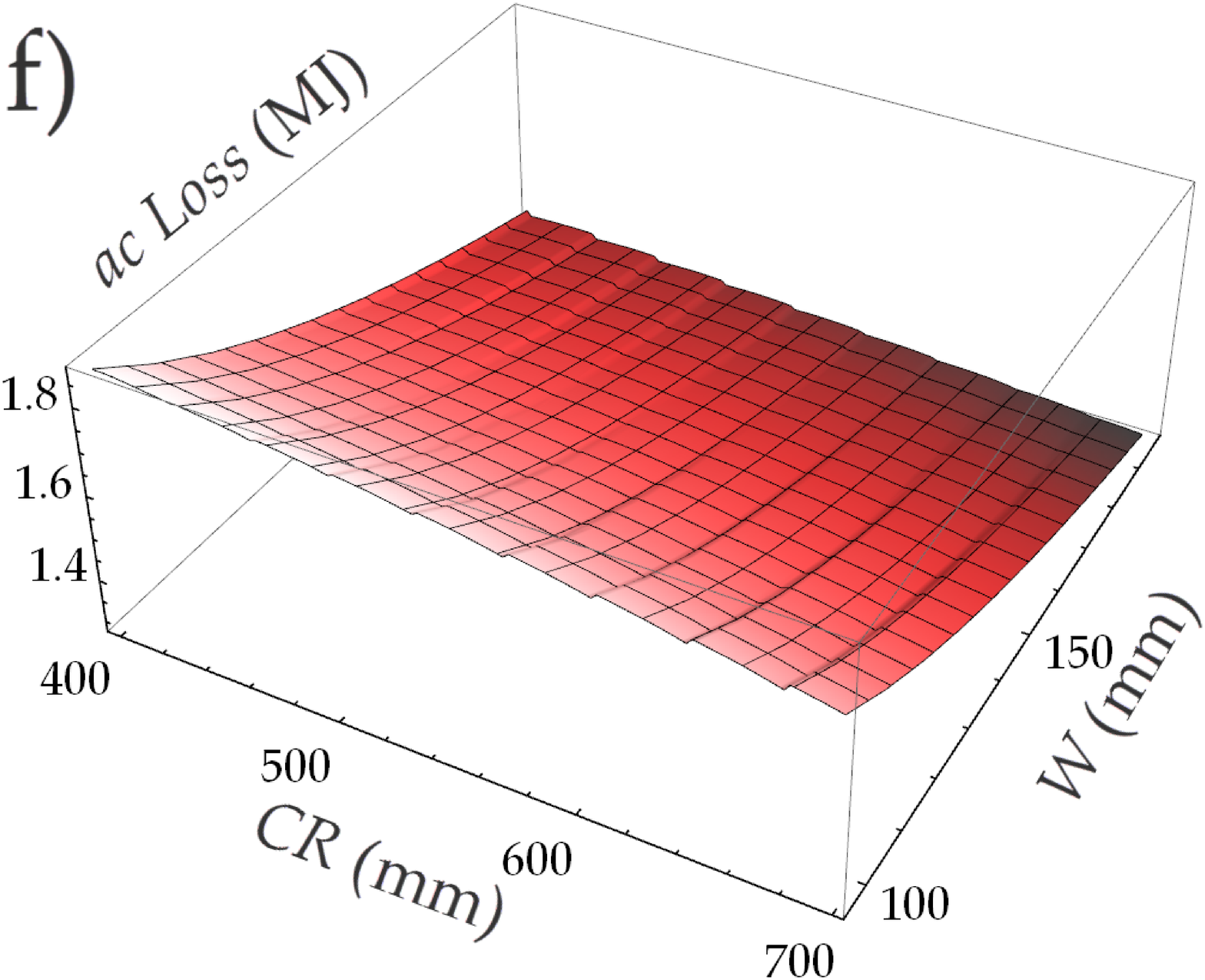}
\vspace*{0.2cm}
\includegraphics[scale = 0.14]{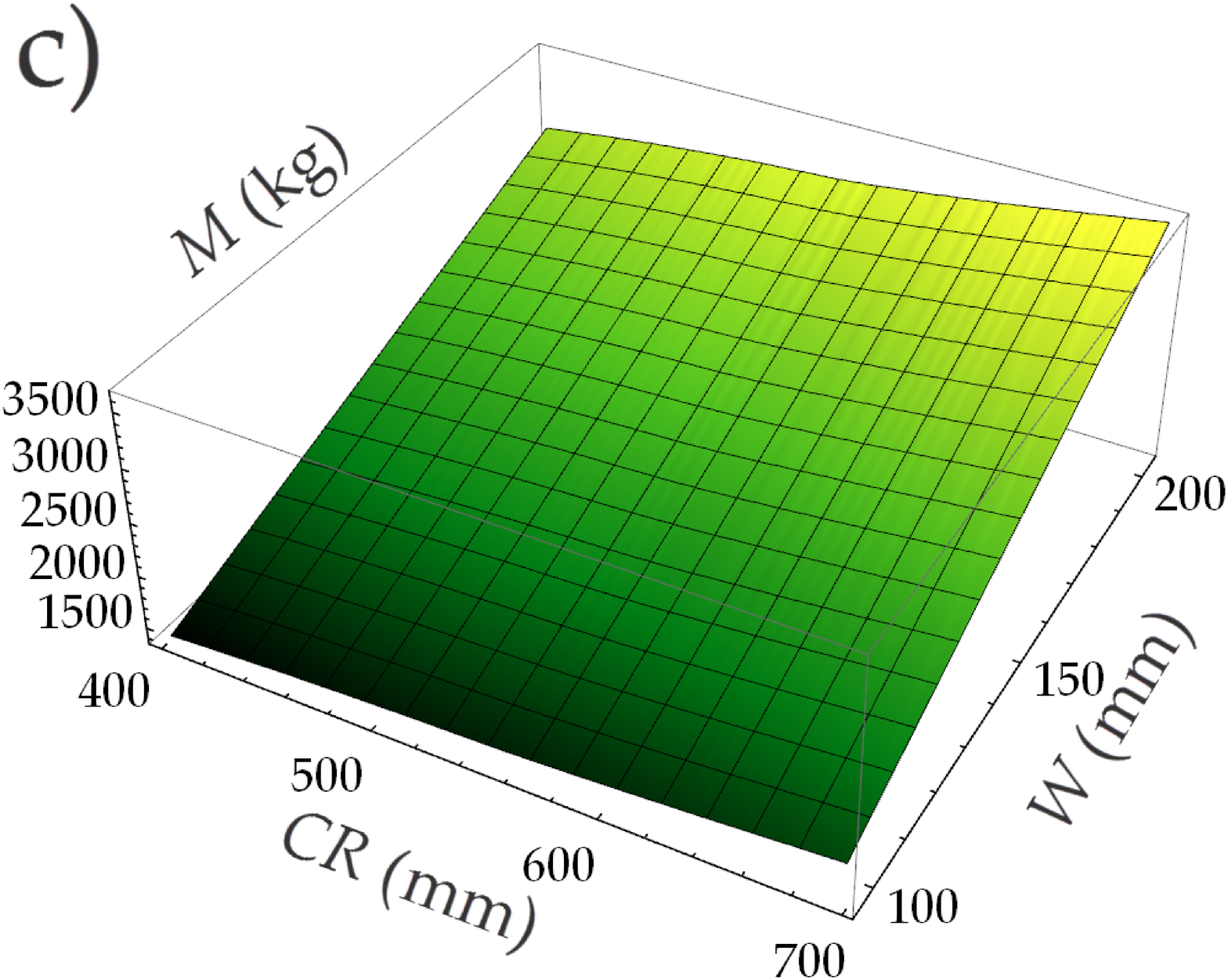}\hspace{2cm}
\includegraphics[scale = 0.14]{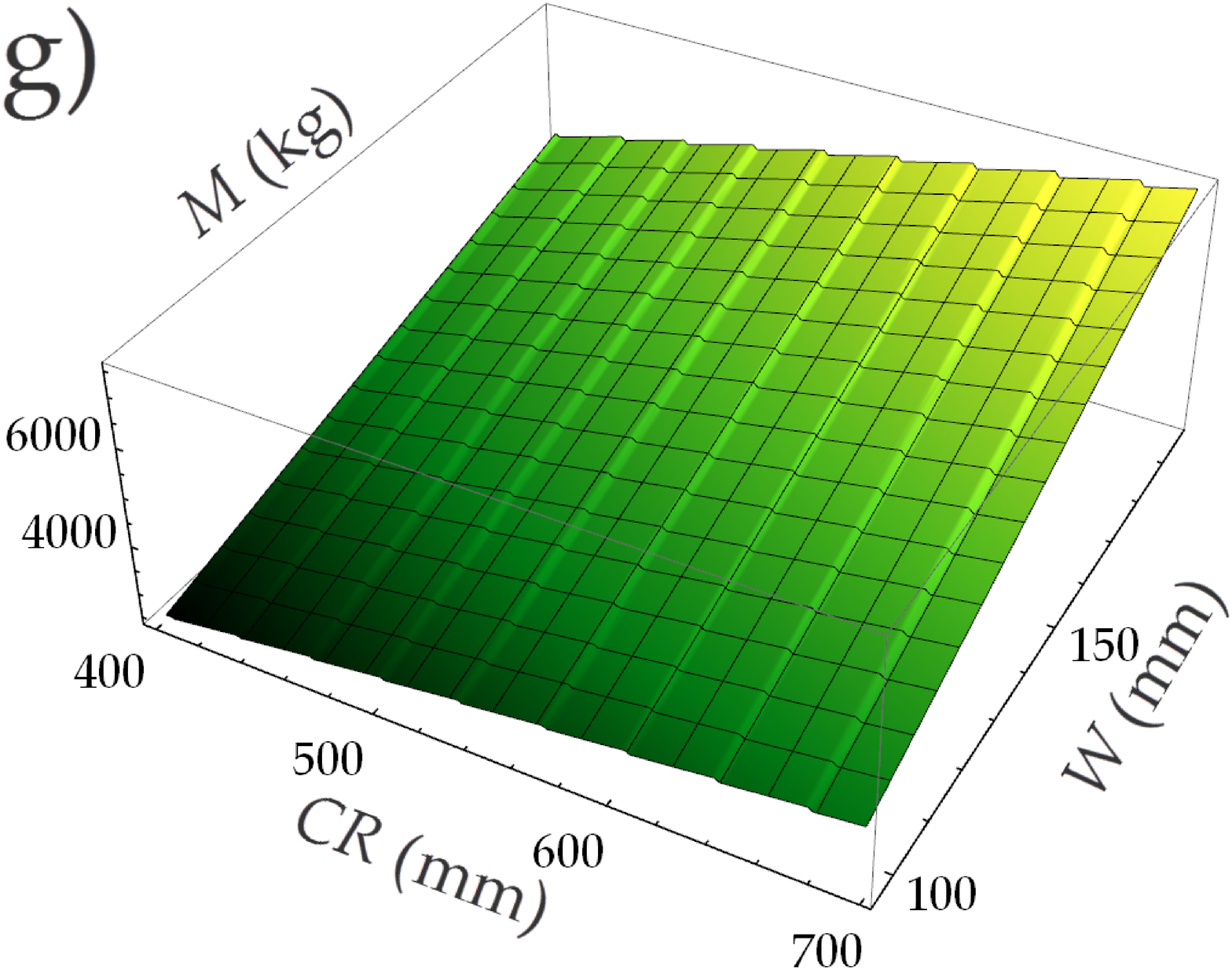}
\vspace*{0.2cm}
\includegraphics[scale = 0.14]{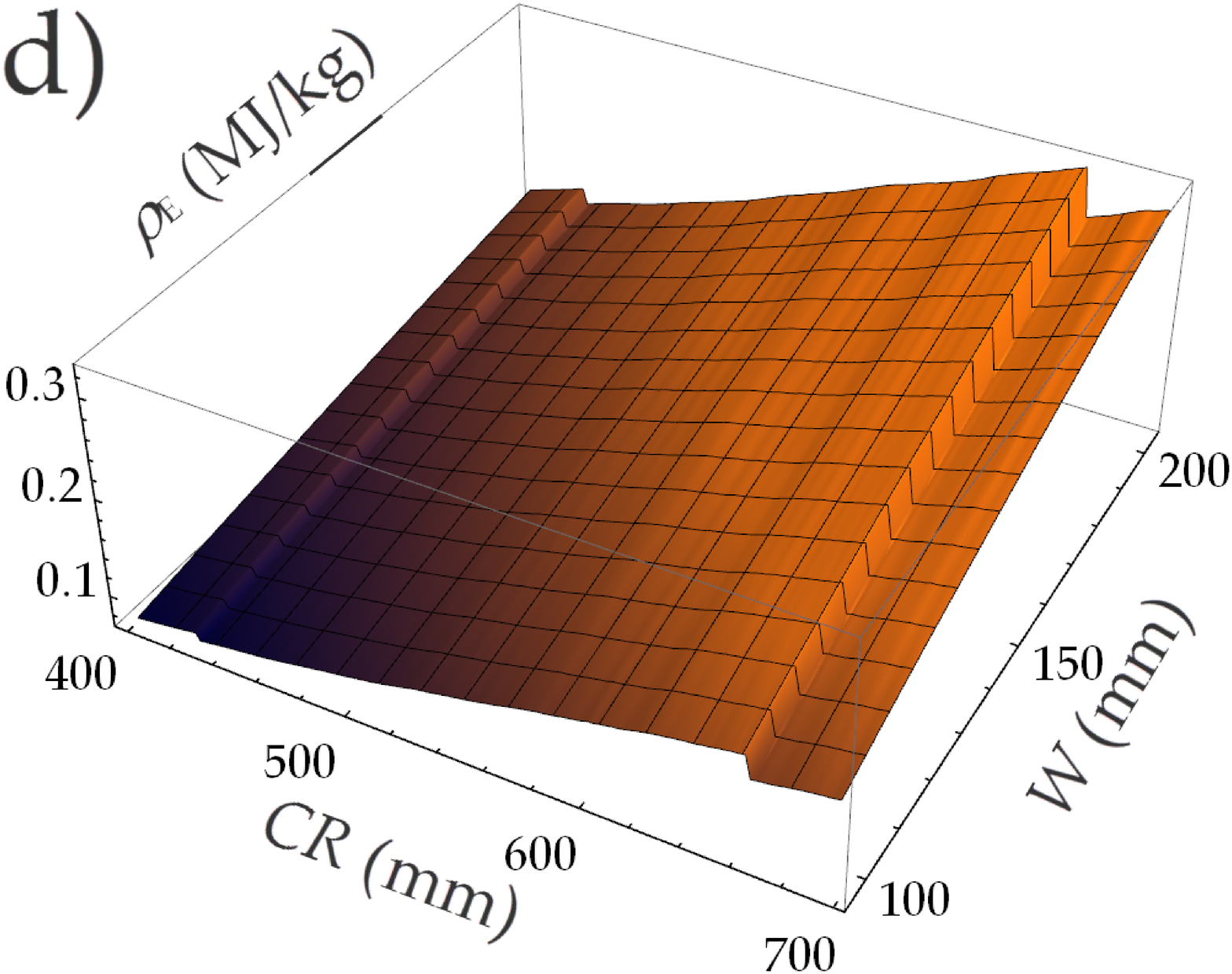}\hspace{2cm}
\includegraphics[scale = 0.14]{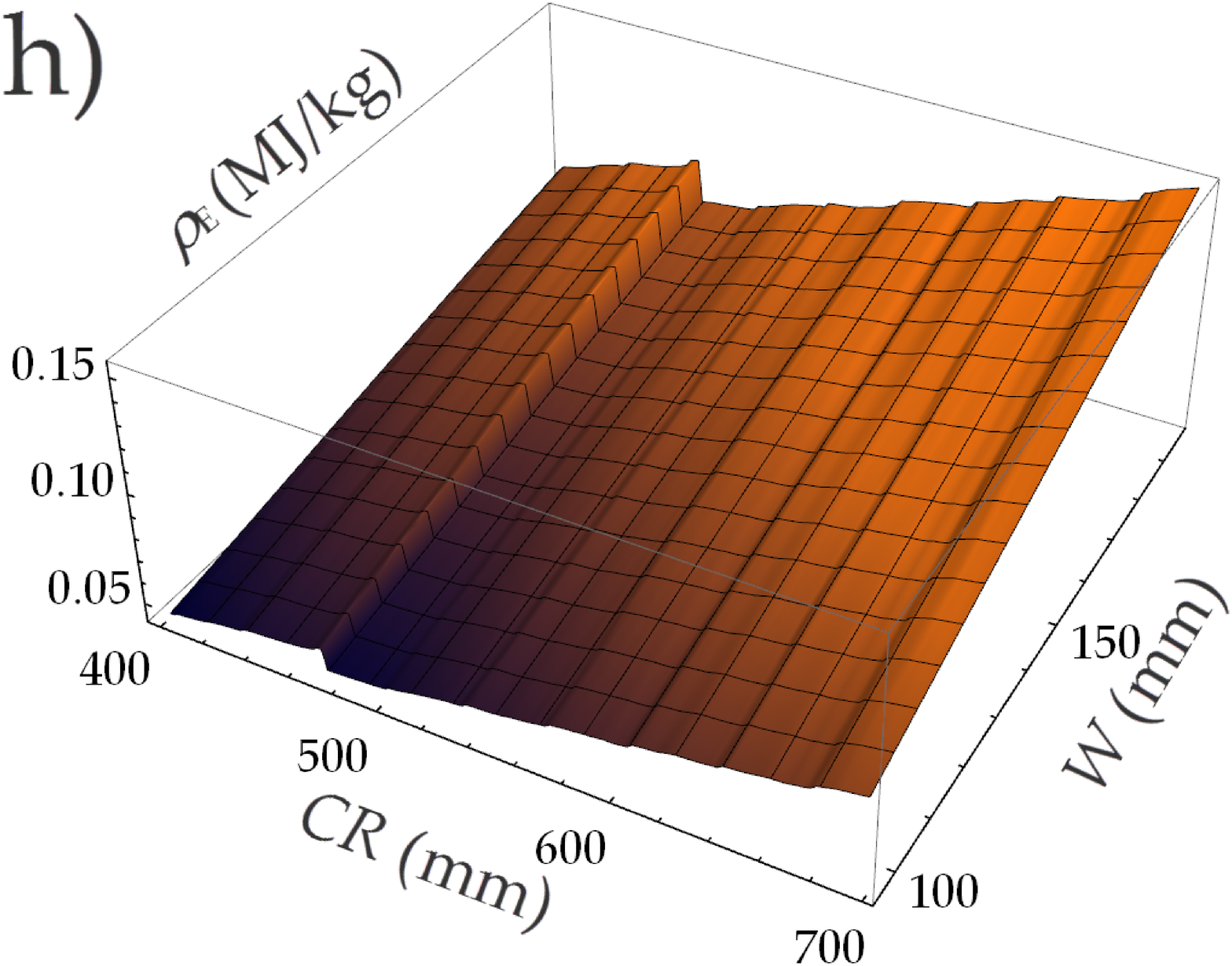}
\caption{Shown are the outputs of the simulation: a) total energy stored, b) ac losses per cycle c) total wire mass, and d) energy density in the YBCO 2G SMES.  The corresponding e) total energy stored, f) ac losses per cycle, g) total wire mass, and h) energy density for MgB$_{2}$ are shown in juxtaposition.  The "ripples" observed in the various quantities are artifacts of the simulation (see text).}  
\end{center}
\end{figure} 

For higher values of $CR$ and $W$, the value of the total energy is actually closer to ${10}^{9}$ J, most likely due to non-linear effects as one can expect a geometric change to bring about.  Larger coils, in radius and width, would have coils in a closer proximity to one another (we never change the values of the $LR$'s once fixed in the beginning), suggesting that fewer field lines will "leak" in the spaces between them, improving on the overall efficiency of the device.  The whole reason why we could not extend the simulation for arbitrarily large $CR$'s and $W$'s had to do with the unphysical overlap of coils that our optimization would bring about.

The values of several SMES parameters were obtained at the highest energy densities from our $CR-W$ configuration space (see Table 2), and from the comparison between the simulations of the 2G YBCO and MgB$_{2}$ based SMES devices, we quickly notice that while the overall dimensions and stored energies are very comparable, a YBCO-based SMES weighs less than half of its MgB$_{2}$ counterpart.  Also, the YBCO device has more than double the energy density compared to the one made of MgB$_{2}$ wire at the present energy-storage capacity.    

\begin{table}[h]
\caption{\label{arttype}Comparison of design parameters for 20-coil YBCO and 32-coil MgB$_{2}$-based SMES devices at maximum energy density.}
\footnotesize\rm
\begin{tabular*}{\textwidth}{@{}l*{15}{@{\extracolsep{0pt plus12pt}}l}}
\br
SMES Type & ${{\rho}_{E}}^{max}$(MJ/kg) & ${M}^{max}$(kg) & ${{E}_{Total}}^{max}$(MJ) & ${CR}^{max}$(mm) & ${W}^{max}$(mm) & ${T}^{max}$(mm)\\
\mr
YBCO & 0.327 & 3,416 & 1,117.07 & 662 & 204 & 68.7\\
MgB$_{2}$ & 0.155 & 7,537 & 1,168.26 & 698 & 196.8 & 50.3\\
\br
\end{tabular*}
\end{table}   

An interesting qualitative difference between the two systems could be observed in their ac losses (Fig. 7b and 7f).  ac losses are proportional to both the magnetic field ($B$ and ${B}^{\perp ab}$ for MgB$_{2}$ and YBCO, respectively), and ${I}_{c}$.  Consequently, we would expect that analytical solutions for ac losses in a SMES would be virtually impossible to obtain, which shows why realistic simulations are truly indispensable in order to assess ac losses and their precise geometrical functional forms.  Here, we see that while ac losses in MgB$_{2}$ decrease monotonically with $CR$ and $W$ (Fig. 7b), in the case of YBCO they increase until reaching a maximum (Fig 7f). 

The critical current decreases with an increase in $B$, suggesting that the evolution of ${E}_{L}$ would depend on the balance of the two.  The wire in a coil carries transport current and simultaneously experiences an applied magnetic field generated by the neighboring turns and from the rest of the SMES, thus ac losses in a coil are expected to be a combination of both transport and magnetization losses \cite{Nguyen}.  An example of the complexity of ac losses could be observed in the behavior of the ac losses as a function of coil dimensions (Fig. 7b and 7f).  While in the case of MgB$_{2}$ the losses decrease monotonically as a function of $CR$ and $W$ (Fig. 7f), the opposite is observed in the case of YBCO, and furthermore, a maximum is seen to form in the $CR - W$ plane.  The conserved quantity in our simulation is the load factor and not energy, which increases with the coil dimensions.  That would mean that to first order we should expect comparable transport losses in the $CR-W$ plane.  However, the reduced proximity of the coils in the geometric space would lead to reduction in the inductive losses, which in the YBCO case likely dominate the ac loss contribution.  Therefore, identifying ac loss maxima is particularly useful for 2G-based SMES devices, particularly for cases where the large radius is fixed. 
 
We can conclude from observing the simulations results that ac losses for large energy storage devices (${10}^{8} - {10}^{9}$ J and beyond) are quite irrelevant.  Notice that the highest ${E}_{L}$'s observed are smaller than 0.5 MJ and 2 MJ per cycle for YBCO and MgB$_{2}$ (see Fig 7 and 8), respectively, whereas the lowest stored energy is 100 MJ for both SMES varieties.  The ac loss is approximately given by the ratio between filament size and coil radius.  Since these are large coils ($\sim$1 m), and filament is 12 mm, we would expect an ac loss on the order of 1$\%$, which is roughly what the calculation shows.  The toroidal geometry may reduce this figure ever so slightly.

However, we note that for devices that store lower energies ($\leq {10}^{7}$ J) ac losses will tend to have a very profound effect, and they must be duly subtracted from the stored energy values for proper accounting of device efficiency.  Given the complicated dependencies of ac losses on device geometry, we can see how at lower stored energies ($\leq {10}^{7}$ J), ${\rho}_{E}$ will most likely exhibit a non-monotonic behavior in the $CR-W$ plane, where our \emph{Radia}-based algorithm will have an even greater relevance in locating design "sweet spots".             

\begin{figure}
\begin{center}
\includegraphics[scale = 0.14]{Fig7YBCONew.eps}\hspace{6cm}
\includegraphics[scale = 0.14]{Fig7MgB2New.eps}\\
\includegraphics[scale = 0.18]{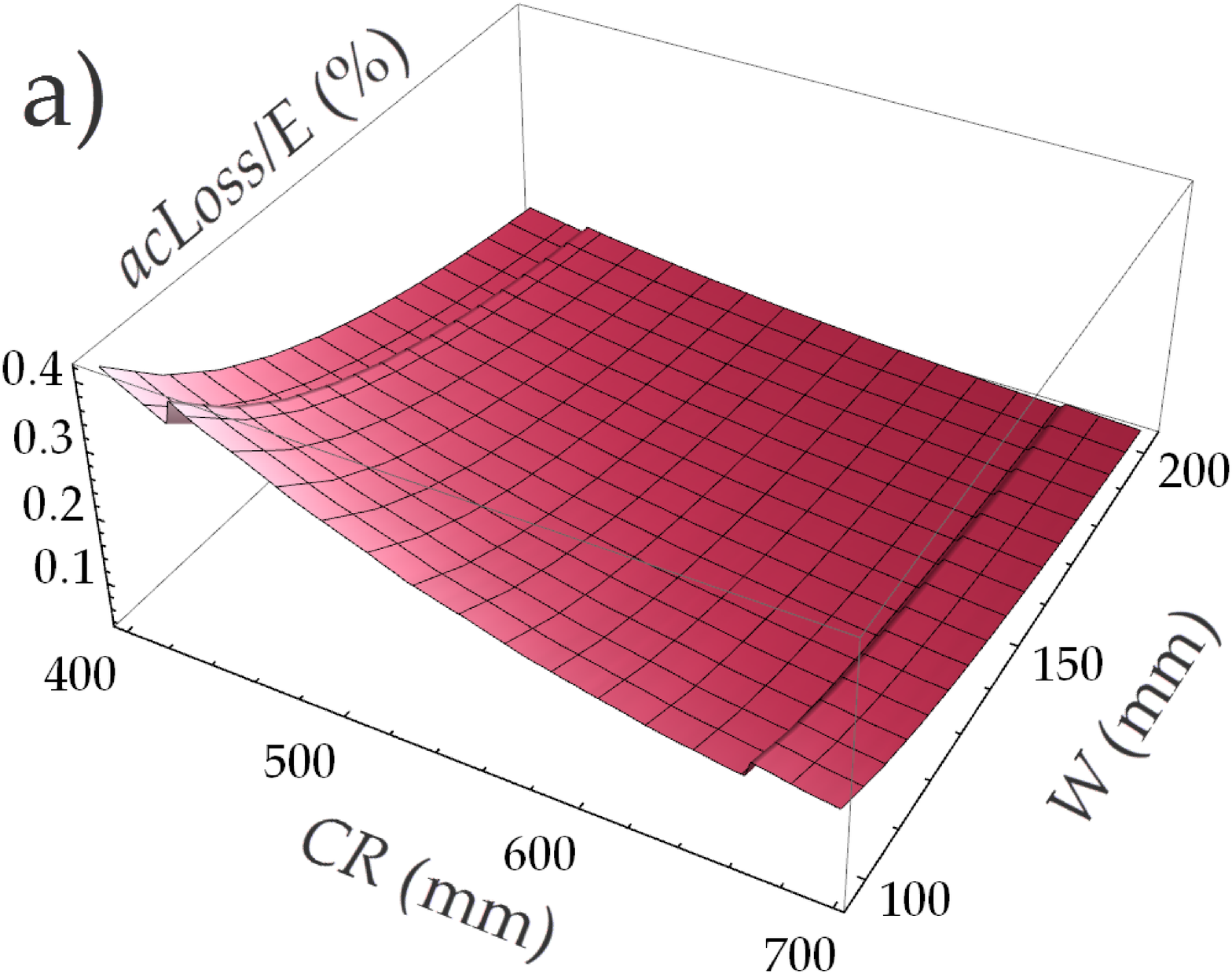}\hspace{2cm}
\includegraphics[scale = 0.18]{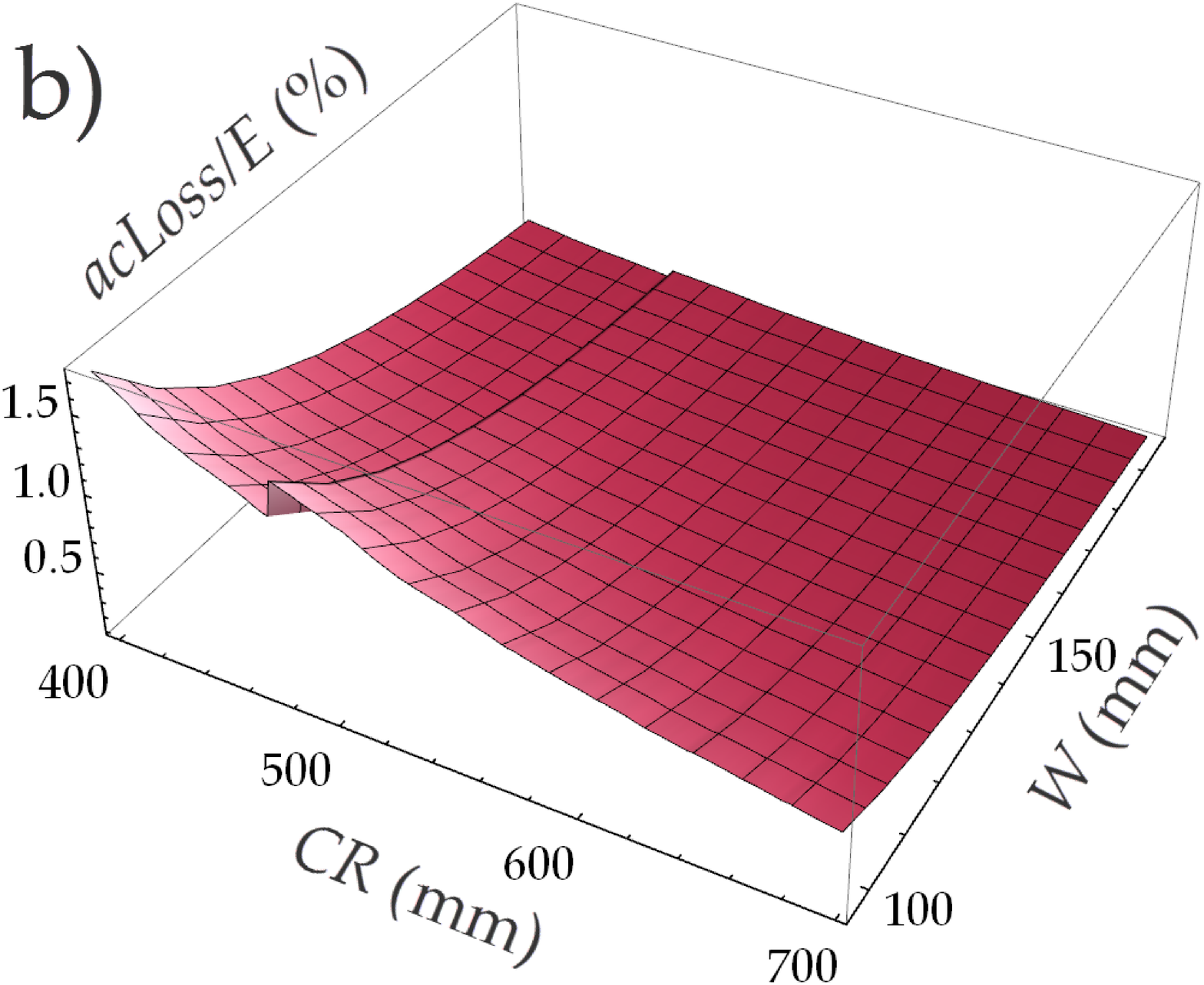}
\vspace*{0.2cm}
\caption{ac losses expressed as percentile of total stored energy for a) YBCO and b) MgB$_{2}$ devices.}  
\end{center}
\end{figure}

Finally, we would like to address the cost aspect of the two types of SMES devices we are considering.  After optimizing each of them at every point in configuration space, we were able to get an estimate of the overall length of wire needed (see Fig. 9).  One thing to notice is that the MgB$_{2}$-based device would require significantly more wire, in absolute length, for comparable amount of stored energy when compared to its YBCO counterpart (see Table 3).
      
\begin{figure}[ht]
\begin{center}
\includegraphics[scale = 0.14]{Fig7YBCONew.eps}\hspace{6cm}
\includegraphics[scale = 0.14]{Fig7MgB2New.eps}\\
\includegraphics[scale = 0.14]{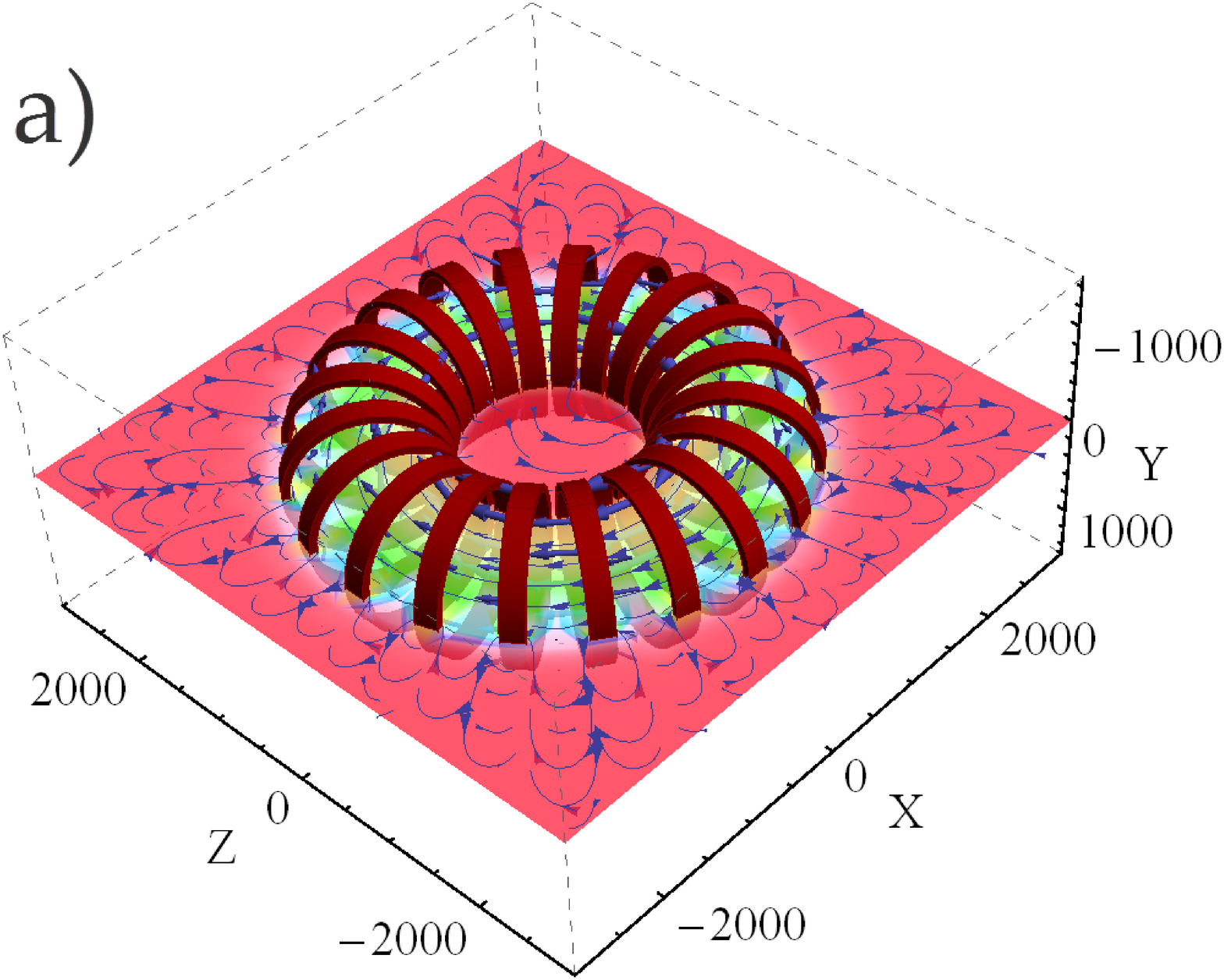}\hspace{2cm}
\includegraphics[scale = 0.14]{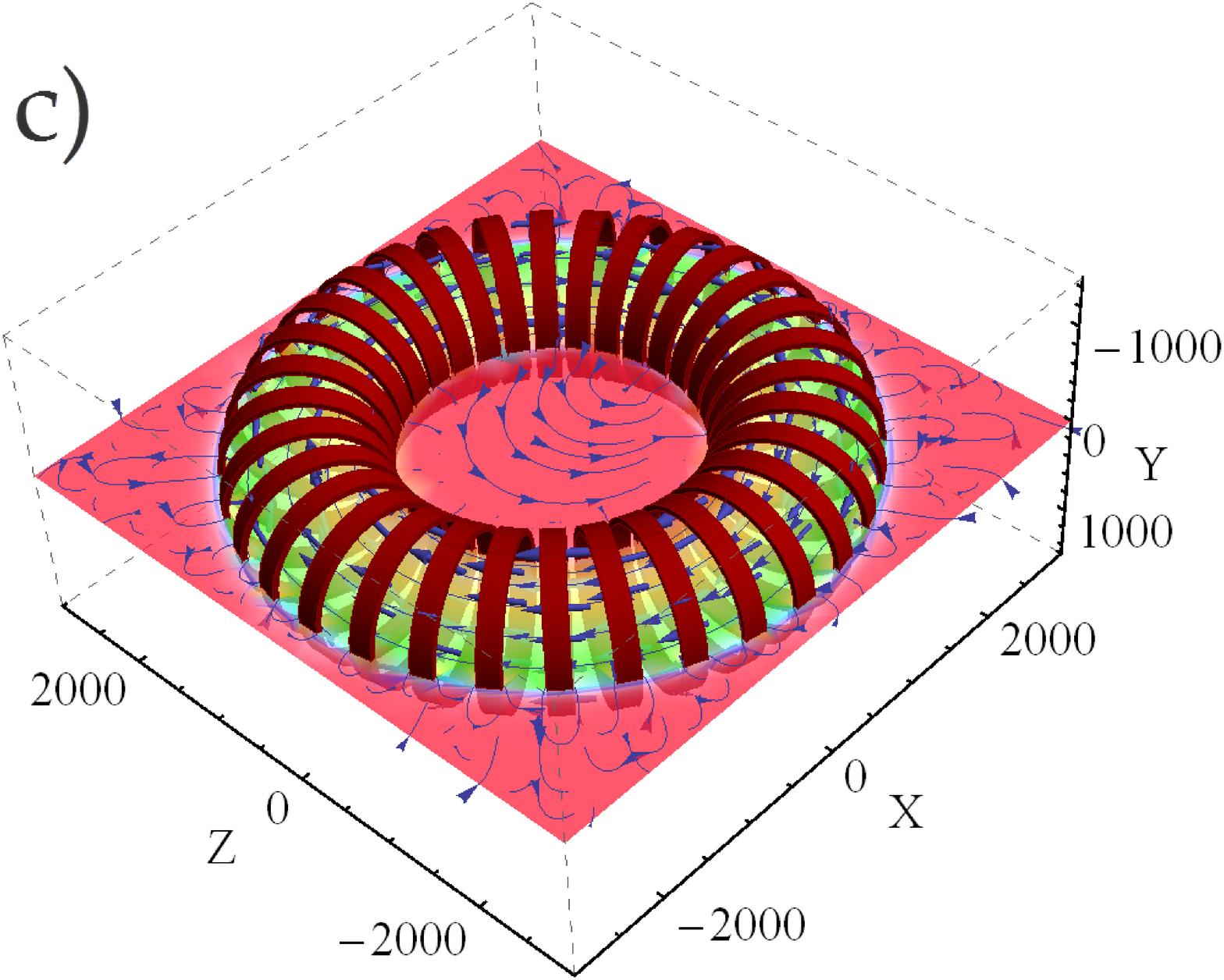}
\vspace*{0.2cm}
\includegraphics[scale = 0.16]{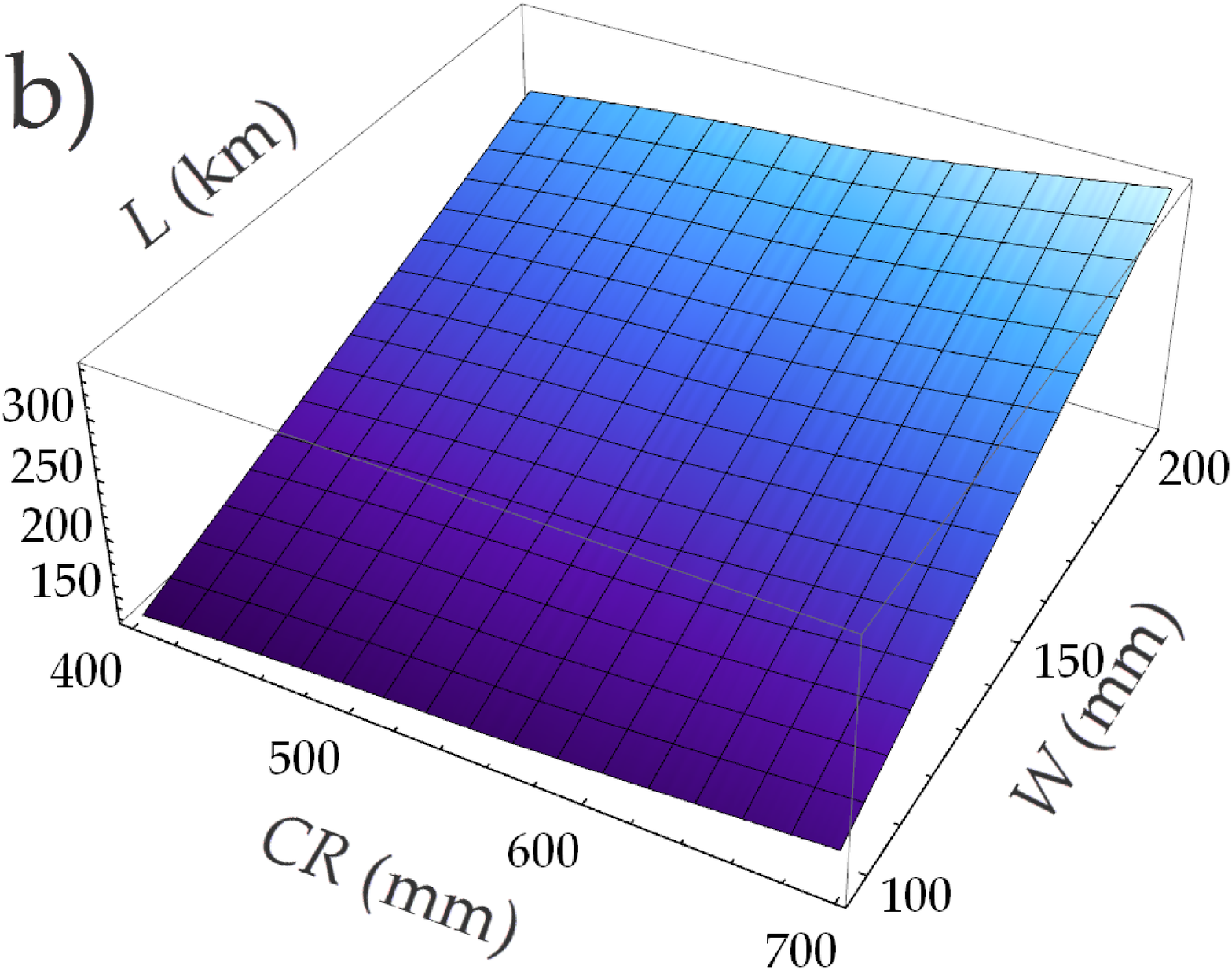}\hspace{2cm}
\includegraphics[scale = 0.16]{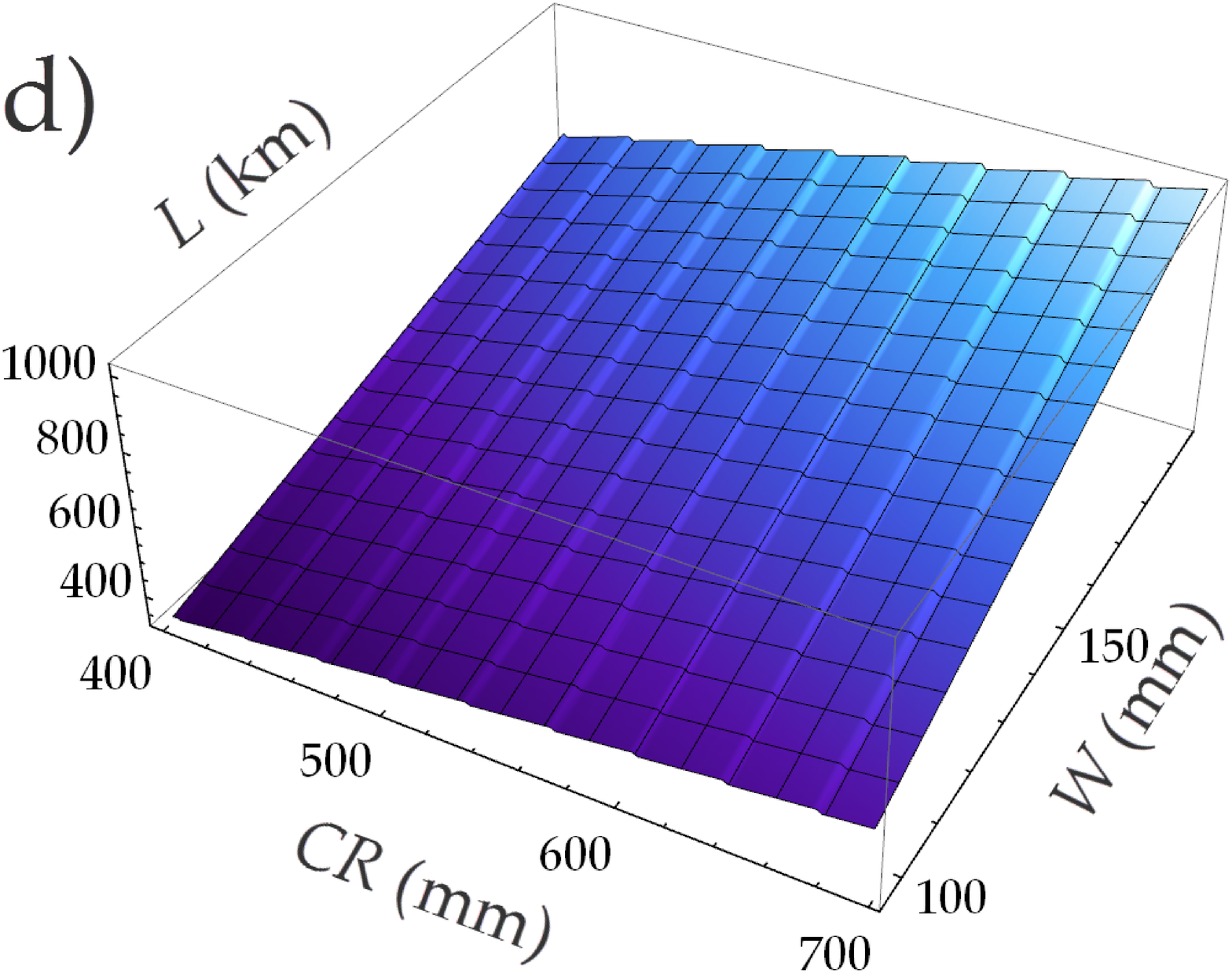}
\caption{Shown are the final configuration outputs at maximum energy density for a) YBCO and b) MgB$_{2}$ devices.  All the dimensions in a) and c) are given in millimeters.  The total wire length requirements for each optimized configuration in $CR-W$ are shown for YBCO and MgB$_{2}$-based devices in b) and d), respectively.}  
\end{center}
\end{figure} 

\begin{table}[h]
\caption{\label{arttype}Comparison of wire length requirements (in kilometers) for 20-coil YBCO and 32-coil MgB$_{2}$-based SMES devices at \emph{i}) $CR$ = 400 mm, $W$ = 100 mm and \emph{ii}) $L$($CR$ = 700 mm, $W$ = 200 mm.}
\footnotesize\rm
\begin{tabular*}{\textwidth}{@{}l*{15}{@{\extracolsep{0pt plus12pt}}l}}
\br
SMES Type & $L$($CR$ = 400 mm, $W$ = 100 mm)(km) & $L$($CR$ = 700 mm, $W$ = 200 mm) (km)\\
\mr
YBCO & 130 & 341\\
MgB$_{2}$ & 332 & 1,018\\
\br
\end{tabular*}
\end{table}   

Therefore, in order to build a 2G wire-based SMES intended to store 100 MJ of energy, on the order of several millions of dollars need to be allocated for 2G wire purchase at today's market prices.  In terms of the sheer benefit offered by reduced weight, and amount of materials used, 2G wire is currently the obvious choice for large-stored energy, large energy density SMES device.

\section{Conclusions}
 
In this paper we demonstrated the viability of \emph{Radia} as a CPU-efficient semi-analytical method for optimizing prospective superconducting energy-storage devices.  By altering various device parameters, such as coil radius and thickness, we were able to calculate the optimal coil thickness and simulate total stored energy and ac losses in a YBCO and MgB$_{2}$-based devices.  

\section{Acknowledgements}
This work was primarily supported by the US Department of Energy, Office of Basic Energy Science, Materials Sciences and Engineering Division, under contract no. DEAC0298CH10886.  Additional support from Air Force Research Lab (V.F.S. and I.K.D.) and NYSERDA (V.F.S.) are also acknowledged.  This effort was also supported by the System Evaluation Division of the Institute for Defense Analyses.  I.K.D. wishes to thank Steve Warner for interest taken in the current work as well as for critical reading of the manuscript, and Michael Ambroso for technical assistance.


\begin{thebibliography}{9}
\bibitem{Ivo4} For the purposes of the YBCO-based SMES calculation we take that the YBCO tape comprises a 2 $\mu$m YBCO,  50 $\mu$m copper, 45 $\mu$m stainless steel and 50 $\mu$m kapton  layers, respectively (see Section 3.3).  We take the tape width to be 12 mm, and the densities of YBCO, copper, stainless steel and kapton to be 6.3$\times$10$^{-3}$ g/mm$^{3}$, 8.92$\times$10$^{-3}$ g/mm$^{3}$, 8.03$\times$10$^{-3}$ g/mm$^{3}$ and 1.42$\times$10$^{-6}$ g/mm$^{3}$, respectively
\end{thebibliography}
\end{document}